\documentclass[smallextended]{article}       % onecolumn (second format)
\usepackage{graphicx}
\usepackage{xcolor}
\usepackage{amsmath}
\usepackage{siunitx}
\usepackage{algorithmic}
\usepackage{algorithm}
\newcommand{\no}{\nonumber}
\newcommand{\halb}{\frac{1}{2}}
\newcommand{\eps}{\epsilon}
\newcommand{\sqrho}{\sqrt{\rho} }
\newcommand{\change}[1]{{ #1}}
\begin{document}

\title{Pressure-tight and non-stiff volume penalization for compressible flows\\

	{\small An immersed boundary method with good conservation properties}}

\author{Julius Reiss\thanks{Technische Universität Berlin, 10623 Berlin,reiss@tnt.tu-berlin.de}}

%	\thanks{  Technische Universität Berlin,       10623 Berlin,  

%		{reiss@tnt.tu-berlin.de}           %  \\

%	}

%}

\maketitle

\begin{abstract}
Embedding geometries in structured grids allows a simple treatment of complex objects in fluid simulations.  
Various methods for embedding geometries are available. 
The commonly used Brinkman-volume-penalization models geometries as porous media, and  approximates a solid object in the limit of vanishing porosity.  
In its simplest form, the momentum equations are augmented by a term penalizing the fluid velocity, 
yielding good results in many applications. 
However, it induces numerical stiffness, especially if  high-pressure gradients need to be balanced. 

Here, we focus on the effect of the reduced effective volume (commonly called porosity) of the porous medium. 
An approach is derived, which  allows reducing the flux through objects to practically zero with little  increase of  numerical stiffness. 
Also, non-slip boundary conditions and adiabatic boundary conditions are easily constructed.   
The porosity terms allow keeping the skew symmetry of the underlying numerical scheme,   by which the numerical stability is improved. 
Furthermore, very good conservation of mass and energy in the non-penalized domain can be achieved, 
for which the boundary smoothing introduces a small ambiguity in its definition. 
 The scheme is tested for acoustic scenarios, for near incompressible  and strongly compressible flows.

\emph{CFD \and  immersed boundaries \and  penalization \and energy stable methods \and   skew symmetry \and      conservation \and   compressible flows \and slip boundaries \and adaptive filter }
 %\subclass{ 76M20 \and 76N05 \and 76S05 }
\end{abstract}

\section{Introduction}

Most flows in technical applications are defined by the geometrical properties of the enclosing or the contained objects, which are often of complex shape.   
To simulate flows in such settings, commonly  grids  are constructed, where   cell boundaries   or grid lines are aligned with the boundaries of the objects. 
This simplifies the implementation of boundary conditions, as these conditions can be enforced at specific cell boundaries or grid lines.   
However, the construction of such grids often is a considerable effort on its own.  

Alternatively, simple Euclidean grids can be used and the object boundaries can be approximated by force-like terms. 
Different approaches are in use. 
\change{
With the immersed boundary (IB) methods the boundary is given by a set of point forces, as in a pioneering work by Peskin \cite{Peskin1972}
for the  simulation of heart valves. 
An IB Method for compressible flow is presented for example by Palma et al. \cite{PalmaTullioPascazioNapolitano2006}.}

The point forces are distributed to the closest grid points by utilizing discrete delta functions.   
The cut-cell approach is often used for finite volume schemes, where the cells are divided  along the object boundary 
\change{ as introduced by Berger et al. \cite{BergerLeVeque1989}, see also  \cite{UdaykumarShyyRao1996,Quirk1994}}. While in principle simple, extra care is needed  to avoid very small or distorted cells.   
Often immersed boundaries are used for fluid-structure  interaction, a recent review is provided by \cite{KimChoi2019}.
An interesting approach was recently  presented in \cite{KhaliliLarssonMueller2019}, in which 
 ghost points and interpolation is combined with summation-by-parts finite differences. 
 The latter guarantees  well-defined fluxes, which is helpful for stability.

This article focuses on \change{Brinkman} penalization, for which not only the boundary but the whole interior of the object is forced. 
This can be viewed as a sponge-like or porous medium, which becomes impenetrable for a vanishing porosity. 
Often the porosity associated force terms do not jump at the interface, instead, they are smoothed over a few grid points to avoid numerical problems. 
It is also advantageous for moving objects, where otherwise the discrete inclusion of grid points yields high-frequency fluctuations in the numerical solution.  

In the most simple case, the porous object is modeled by the Darcy term, an extra friction term punishing a flow velocity relative to the object's movement. 
\change{Its original application is to describe flows through porous media in technical or geological applications, as for example done by Masson and co-workers  \cite{EymardGuichardHerbinMasson2012}.
As a penalization method it 
 is used by Boiron et al.} 
\cite{BoironChiavassaDonat2009} for compressible flows and \change{ by Farge et al.} \cite{SchneiderFarge2005} for incompressible flows. 
\change{Angot et al. present} \cite{AngotBruneauFabrie1999} convergence theorems to the true solution for incompressible flows. 
Despite its simplicity, it often yields good results. 
For example, a highly resolved incompressible flapping flight simulation, as a flow around moving objects, is presented \change{by Engels et al.} \cite{EngelsKolomenskiySchneiderSesterhenn2016}. 
\change{Due to}  the finite strength of the Darcy term, the non-slip condition is effectively replaced by an exponential decay of the flow velocity in the object. 
The smearing of the interface increases this effect. 
This can be mitigated by a locally refined grid of a multi-resolution approach, which is, 
 combined with an artificial compressibility method, presented by \change{Engels et al.} \cite{EngelsSchneiderReissFarge2019}.
The effect of the smoothing of the mask function on the convergence of the solution is investigated for incompressible flows \change{by Hester et al.} \cite{HesterVasilBurns2020}. 
It is found that the slow convergence observed in earlier work is due to a displacement of the effective boundary, which can be improved by 
an optimally  chosen smoothing.

%\cite{PelantiShyue2019} 
%\cite{SaurelAbgrall1999}

\change{Liu and Vasilyev  \cite{LiuVasilyev2007}  treat} compressible flows by a Brinkman penalization method popularizing the approach in the compressible regime. 
They include both, the linear friction relative to the object (Darcy's law), and the effective volume fraction remaining for the fluid $\phi$,
which is often called \emph{porosity}. 
In this publication, it will be referred to as volume fraction to distinguish it from porosity as a complex physical setting.   
The difference to the here proposed scheme will be discussed below in Sec.~\ref{sec:eqn}.  
\change{Komatsu et al.  \cite{KomatsuIwakamiHattori2016} calculate} acoustic radiation of flows passing different objects by 
a variant of the method of Liu and Vasilyev \cite{LiuVasilyev2007}, which is modified to be invariant under Galilean transformation. 

The shallow water equations are mathematically very similar to compressible flows. 
In the work of Kevlahan et al. \cite{KevlahanDubosAechtner2015}, a Brinkman penalization method is derived for these equations from a variational principle, paying particular attention to the volume fraction.  
It is pointed out that the transport of momentum and mass should be consistent, and that the speed of sound should not be modified by the volume fraction, to avoid stiffening of the equations. 

Following the discussion of Kevlahan et al.~\cite{KevlahanDubosAechtner2015}, the effect of the volume fraction is investigated in this report. 
Similar to the shallow water equations presented there, compressible Navier-Stokes equations are derived, where consistency in the treatment of the volume fraction is emphasized. 
In this publication, it is shown that the volume fraction is helpful \change{in the context of Brinkman penalization} in many aspects, namely in avoiding stiffness, in creating slip boundaries, and good adiabatic boundaries, and for conservation properties. 
Interestingly the derived equations were presented before by Liu and Vasilyev \cite{LiuVasilyev2007}, 
\change{ as the fundamental equations for flow through porous material by Darcy.}  
but where turned down following arguments in Nield and Bejan \cite{NieldBejan2006} and Beck \cite{Beck1972}, suggesting structural problems.
However, here no such structural problems appear, instead, it is found that the volume fraction and Darcy term can be chosen largely independently, resulting in a substantial extension of possible boundary conditions
\change{ for the Brinkman penalization method. 
	
Recently Kemm et al. \cite{KemmGaburroTheinDumbser2020} described a similar method derived from a two-phase flow based on the Baer-Nunziato model \cite{BaerNunziato1986}, reducing it to one fluid phase and prescribing the front movement. 
It is referred to as a diffuse interface approach and not as Brinkman penalization. 
These equations agree with the here obtained equations when the Darcy friction terms are neglected. 
Source terms similar to the Darcy friction are discussed for the Baer-Nunziato model by Kapila et al. \cite{KapilaMenikoffBdzilSonStewart2001}. 
While here the focus here the is on static boundaries, Kemm et al. \cite{KemmGaburroTheinDumbser2020} successfully investigates moving boundaries and discusses in a rigorous manner  
the non-penetration condition. 
Their numerical implementation builds on the ADER predictor-corrector discontinuous Galerkin approach by Dumbser et al. \cite{BustoChiocchettiDumbserGaburroPeshkov2020,DUMBSER20088209,DUMBSER20083971}, where the fluxes are calculated by an approximate Riemann solver.
The Riemannn problem for the Baer-Nunziato model is discussed in depth by Andrianov et al. \cite{AndrianovSaurelWarnecke2003,AndrianovWarnecke2004} and Han et al. \cite{HanHantkeWarnecke2012}. 
A Riemann solver for the Baer-Nunziato model are discussed for example by Torro et al. \cite{TokarevaToro2010} (HLLC-type) or Pelanti et al. \cite{PelantiBouchutMangeney2008} (Roe type).}

\change{
The perspective of a two-fluid model allows an elegant description of fluid-structure interaction by an Euerian description of the elastic solid as described by Favrie et al. \cite{FavrieGavrilyukSaurel2009,NdanouFavrieGavrilyuk2015} and including 
reactive matter as describe  by Michael et al. \cite{MichaelNikiforakis2018}. 
Acoustic waves in geological settings for non-simple geometries are discussed by Taveli et al. \cite{TavelliDumbserCharrierRannabauerWeinzierlBader2019}. 
Similarly, free surfaces of fluids are described by Dumbser \cite{Dumbser2013} and by Gaburro et al. \cite{GaburroCastroDumbser2018}.}

\change{
	The basic principle of the here presented method is closely related to the one of Kemm et al. \cite{KemmGaburroTheinDumbser2020}. 
It is motivated by the Brinkman penalization method and not by the Baer-Nunziato model, 
thereby bringing these different modeling approaches together. 
The simple geometry treatment allows to use (conservative) finite differences, which are fully explicit 
and straightforward to implement. 
To permit the application of finite difference to a wide range of problems a filtering strategy must be adapted, which allows a locally varying filter strength and is conservative for a variable reduced volume $\phi$. 
The resulting method is efficient and is tested on acoustic, strongly compressible, and near incompressible flows. 
}

The paper is organized as follows. 
In section \ref{sec:eqn} the basic equations are derived following the ideas described in \cite{KevlahanDubosAechtner2015} and
the conservation properties are discussed. 
The application of a filter demands a modification, which is described in section \ref{sec:filter}. 
%Various aspects are discussed with 
Numerical examples underpinning the claims are presented in section \ref{sec:examples}. 
First, the good acoustic properties are discussed in Sec. \ref{sec:ac} including the stiffness of the method and the containment for large pressure gradients. 
On the other hand, in Sec.~\ref{sec:flows} the flow configuration of potential flow with slip around a cylinder and a supersonic flow along a wedge is shown. 
The conservation properties are tested for a strongly unsteady blast wave in containment. 
Finally, a vortex street as an example for a simulation with a refined grid is presented. 
The \change{results} are summarized in sec \ref{sec:conclusion}. 
\change{The appendix~\ref{app:detail} provides technical details of the finite difference method and the filter.}

\section{The penalized equations} 
\label{sec:eqn}
Here, the modification of flow equations to represent immersed objects will be derived in analogy to one-dimensional gas dynamics.

\subsection{Derivation of the analytical equations}

The modified equations are to include the effects of reduced effective volume fraction 
\change{$\phi= V_\mathrm{fluid}/V_\mathrm{total}$}, since the volume of the material occupied by porous media is unavailable for the fluid. 
It is expected that the equations mainly scale with the volume fraction since mass, momentum, and energy scale likewise. 
The Darcy term, describing linear friction proportional to the relative velocity between flow and object, is added in a second step. 
As stated \change{by Kevlahan et al.} \cite{KevlahanDubosAechtner2015} for the shallow water equations, it is important that mass and momentum (and for the Navier-Stokes equations, energy) are transported with the same speed. 
We also want to keep the speed of sound $c$ unaltered by the penalization, since this suggests that the same CFL condition CFL = $ (|u| + c) \Delta t/\Delta x $ stays valid \cite{KevlahanDubosAechtner2015}. 
The validity of this assumption is discussed below in Sec.~\ref{sec:stiff}. 

\begin{figure}[h]
	\begin{center}
		\includegraphics[viewport = 0 0 400 210, clip, width=.42\linewidth]{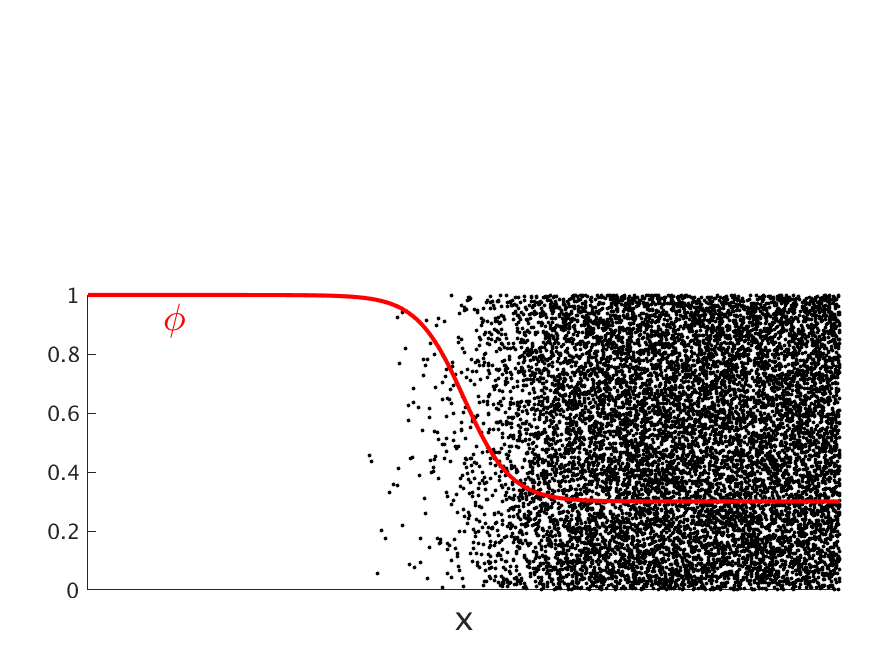}
		\includegraphics[viewport = 0 0 410 210, clip, width=.42\linewidth]{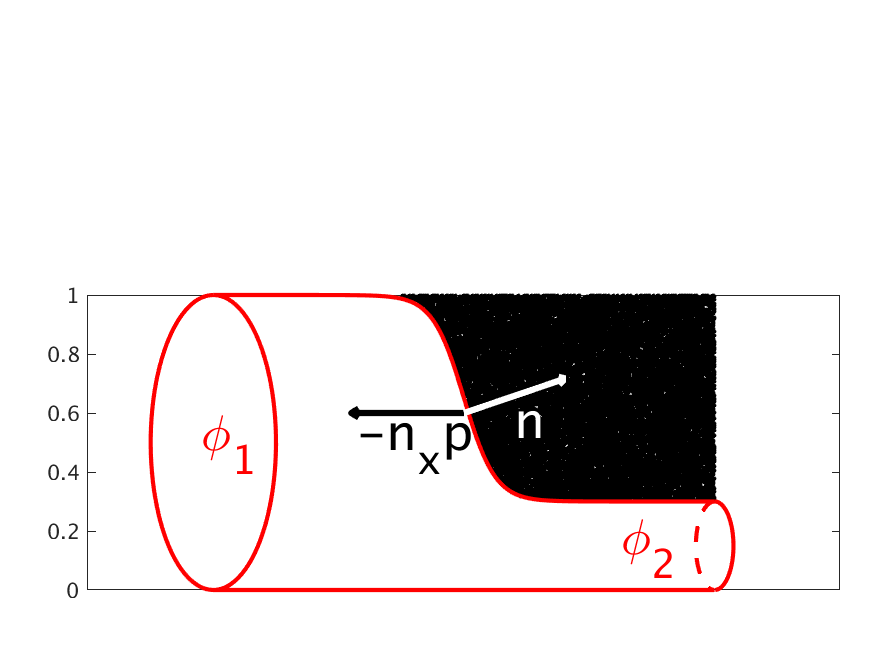}
		
	\end{center}	
	\caption{The effective volume for the fluid can be interpreted in a quasi-one-dimensional flow as a reduced 
		cross-section by reorganizing the material of a flow tube, yielding a configuration known from the theory of stream lines. 
		  This gives an extra pressure source term in the momentum equation. }
	\label{fig:effCross} 	
\end{figure}

Instead of deriving the equations from first principles by the theory of variations as in \cite{KevlahanDubosAechtner2015}, we observe that for one dimension the Euler equations of varying cross-section 
\change{are structurally similar to the equations derived in \cite{KevlahanDubosAechtner2015}. 
	The speed of sound is independent of the cross-section $\phi$, and the transport of mass and momentum are modified in the same way. 
Further, the essential modification by $\phi$ in \cite{KevlahanDubosAechtner2015} is a non-conservative pressure gradient $\phi\partial_x p$,    as it appears in one-dimensional gasdynamics, see eqn.~\eqref{mom}. 
} 
This can be motivated by the fact that no specific distribution of the porous material was prescribed. 
By this, it can be reorganized \change{microscopically} as a tube of reduced cross-section, see Fig.~\ref{fig:effCross}. 
\change{This was recognized before, see \cite{AndrianovWarnecke2004} and references therein.}

This leads to an extra factor $\phi$ in every flux term and an additional source term $p\partial_x \phi$ for the momentum equation, \change{which} 
is detailed in text-books of gasdynamics, e.g. \cite{Anderson1990}. 
\change{A similar modification of the pressure gradient can be identified in the shallow water equation for an uneven bottom height \cite{KevlahanDubosAechtner2015} and the Navier-Stokes equation in cylindrical coordinates, where the 
	role of the reduced cross-section is the circumference proportional to the radius $r|$\footnote{The radial momentum equation can be written as $\partial_t (\rho r u_r ) + \partial_t(r\rho u_r u_r ) + \partial_z (r\rho u_r u_z) + r\partial_r p= 0  $ }. 
	It is also found for porous material \cite{NieldBejan2006} and in the Baer–Nunziato
	model for multi-phase flow \cite{AndrianovWarnecke2004}, see below.
	In all these applications a reduction of a cross-section-like property yields a source term in the momentum equation of this type. }

The equations for mass, momentum, and energy in one dimension without friction are therefore just the equations for one-dimensional gas dynamics in varying cross-section $\phi$ 
\begin{align}
	\change{\phi\partial_t \rho} + \partial_x( \phi \rho u)  &=  0  \label{mass1D} \\ 
	\change{\phi\partial_t (\rho u)} + \partial_x( \phi \rho u u) + \partial_x \phi p &= p\, \partial_x \phi \label{mom1D}\\ 
	\change{\phi\partial_t (\rho e_t ) 
		+
	 \partial_x( \phi \rho u e_t         ) +
   \partial_x(             \phi  u {p} ) }  &=  0.
	\label{energy1D}
\end{align}  

We introduced the mass density $\rho$, the velocity $u$, the pressure $p$ and the total energy 
$e_\mathrm{t} = u^2/2 + e_\mathrm{in}$, where we in the following assume the internal energy of an ideal gas $ e_\mathrm{in} = \frac{p}{\rho}\frac{1}{\gamma-1} $, with the 
adiabatic index $\gamma$. 
In principle, the varying cross-section $\phi$ can have arbitrary (positive) values. 
In the following it is identified with the reduced volume by porosity, so that it is in the range $\phi=[0,1]$.
In the vanishing limit $\phi \to 0 $ no volume is left for the fluid. 
The volume fraction appears linearly in all terms on the left-hand side, whereas the term on the right-hand side is a source term without a derivative of any of the dynamical variables. 
From this, it follows that the equations 
are hyperbolic with unchanged characteristic velocities: the acoustic waves with $\lambda_\pm = u\pm c $ with the speed of sound $c=\sqrt{\gamma p /\rho}$ and the entropy wave $\lambda_s = u$. 
Further, all flux terms are modified in the same manner, 
so that the consistency demand of \cite{KevlahanDubosAechtner2015} is automatically fulfilled. 
If $\phi$ is constant \change{(and non-zero)}, the equations reduce to the standard one-dimensional Euler equations. 

Due to the source term $p\, \partial_x \phi$, the momentum is not conserved, whereas the total mass and energy are conserved.
This is physically sound, since a change in cross-section creates reflections, whereas mass and energy are unaltered.

The equations (\ref{mass1D})-(\ref{energy1D}) can be generalized straightforwardly to multiple dimensions and augmented with dissipative terms.
Summing convention is assumed for all Greek indices $\alpha, \beta =1,2,3$ marking spatial directions. 
\change{The two pressure terms in the momentum equation have been combined 
$\partial_{x_\alpha} (\phi p ) - p\partial_{x_\alpha} \phi   = \phi \partial_{x_\alpha}  p 	 $}, resulting in the equations 
\begin{align}
	\partial_t (\phi \rho) + \partial_{x_\alpha}( \phi \rho u_\alpha)   &=   0   
	\label{mass} \\ 
	\partial_t (\phi \rho u_\alpha) + \partial_{x_\beta}( \phi \rho u_\beta u_\alpha) + \phi \partial_{x_\alpha} p  &=     \phi  \chi  (u_\alpha^\mathrm{t} - u_\alpha  )
	+ \partial_{x_\beta} (\phi \tau_{\alpha \beta}) 
	\label{mom}\\  
	\partial_t (\phi \rho e_t ) + \partial_{x_\alpha}( \phi \rho u_\alpha  e_t   +  \phi  u_\alpha {p} )&=  
	\partial_{x_\alpha} ( \phi u_\beta  \tau_{\alpha\beta} ) + \partial_{x_\alpha} (\phi \lambda \partial_{x_\alpha}  T).
	\label{energy}
\end{align} 
\change{Further}, the Darcy friction term is included in the momentum equation with a spatially dependent force strength $\chi = 1/\eta$ and the target value $ u_\alpha^\mathrm{t} $, which is the speed of
the immersed object, which is always zero in this report. 
We also assume that the objects are static, i.e. $\phi $ and $\chi$ are time independent, in the following. 
\change{The dissipative term} contain the temperature $ T = p/\rho  R W  $ with the universal gas constant $R$ and the molecular weight $W$. 
\change{The viscous friction is given by} 
$\tau_{\alpha\beta  } = \mu  \left(\partial_{x_\beta}  u_\alpha + \partial_{x_\alpha}  u_\beta \right) + (\mu_d   -2/3 \mu)  \delta_{\alpha,\beta}   \partial_{x_\gamma}  u_\gamma $ with $\mu$ and $\mu_d$ the shear and volumetric friction and $\delta_{\alpha,\beta}$ the Kronecker delta. 
Similar to the Darcy term, a penalization can be added to the energy equation to enforce e.g. isothermal boundary conditions, \cite{BoironChiavassaDonat2009}. 
These dissipative fluxes are scaled by $\phi$ since these fluxes take place in the fluid part only. 
It also keeps the symmetry of the dissipation terms, i.e. guarantees negative semi-definite operators. 
%\medskip 

\paragraph{Comparison with former schemes} 
\change{ Liu and Vasilyev \cite{LiuVasilyev2007} discuss different equations for the Brinkman penalization, 
	of which some agree with here proposed equations, while the equations used in the numerical implementation \cite[eqn. (18-20)]{LiuVasilyev2007} do not.} 
For example their mass equation is 
\begin{align}
	\partial_t \rho  = - \left[ 1+ \left( \frac 1 \phi -1\right) \chi\right] \partial_\alpha m_\alpha 
\end{align}
 with $\chi$ the mask function and $m_\alpha = \rho u_\alpha$. 
 \change{In contrast,} the mass equation in the initial discussion \cite[(eqn 2)]{LiuVasilyev2007} and the momentum equation \cite[(eqn 5)]{LiuVasilyev2007}, attributed to Wooding \cite{Wooding1957}, agrees with our equations\footnote{Observe the Dupuit–Forchheimer relationship $v_\alpha = \phi u_\alpha$, used there.}. For the latter, the mass equation (\ref{mass}) needs to be split off the momentum equation (\ref{mom}) 
 \begin{align}
  \rho	\partial_t ( u_\alpha) + \rho u_\beta  \partial_{x_\beta}(u_\alpha) +  \partial_{x_\alpha} p  &=      \chi  (u_\alpha^\mathrm{t} - u_\alpha  )
 	+ \frac 1 \phi \partial_{x_\beta} (\phi \tau_{\alpha \beta}) 
 	\label{momCon}
 \end{align}
by which the volume fraction cancels in all terms beside the shear friction term. This is structurally also found by Kevlahan et al. \cite{KevlahanDubosAechtner2015}. 
However, the (steady state) momentum equation given by Wooding \cite[(eqn.~6)]{Wooding1957} in the cited reference is 
\begin{align}
	\frac 1 \rho \nabla p - \mathbf{g} + \frac 1 k \nu \mathbf{q} = - \phi^{-2} \mathbf{q}\cdot \nabla \mathbf{q} 
\end{align}
with $\mathbf{q}= \phi \mathbf{u}$ (aligned to our notation), with $\mathbf{g}$ the gravity vector and $k$ the Darcy friction factor. 
The combination of the pressure term and the non-linear transport term disagrees with \eqref{momCon} and not all terms have the same order in $\phi$.

Liu and Vasilyev abandon Wooding's equation, \change{similar to the here proposed}, due to an argument of Beck \cite{Beck1972}, who, however, uses a Darcy equation (his equation 1), where a constant $\phi$ does not cancel. Beck finds an insufficient number of boundary conditions for his buoyancy flow, however, no problems are apparent for the \change{here used} equations (\ref{mass}-\ref{mom}). 
The momentum equation of Nield and Bejan \cite[(page 8)]{NieldBejan2006} agrees with the here found equation. 
They argue to drop the non-linear transport, by citing Beck, and by practical considerations for a real porous material, which does not seem imperative for our artificial application of porous \change{model}. 
Instead, it is shown in the simulations that equations used here, with all terms linear in $\phi$, allow to choose the volume fraction $\phi$ and the Darcy term largely independently, making it possible to model different boundary conditions.  
\change{The way $\phi$ enters the equations in this article agrees with the equations of Kemm et al. \cite{KemmGaburroTheinDumbser2020}, derived from the two-phase flow of Baer-Nunziato model. 
	A similar approach for acoustic waves is given by Tavelli et al.  \cite{TavelliDumbserCharrierRannabauerWeinzierlBader2019}. 
 }

%\medskip 
\change{
\paragraph{Moving Objects} 
Moving objects imply a time dependency of the cross-section $\phi=\phi(t)$, which 
results in additional terms in the governing equations. 
These are derived here for the sake of completeness and fully agree with Kemm et al. \cite{KemmGaburroTheinDumbser2020}. 

A time-dependent reduced volume can be interpreted as a flux term, see Fig.~\ref{fig:effCrossDot}. 
For example for the mass equation, the balance equation of an infinitesimal volume is 
\begin{align}
\int_\Omega (\partial_t \rho) \mathrm{d}V 
+ (\phi  \rho u )_{x_2}  - (\phi  \rho u )_{x_1}    
+ (  \rho v)_{y=\phi}  - (  \rho v)_{y=0}    
= 0 .\label{movingVol}
\end{align}
 Here, we assume that the horizontal fluid velocity is constant for one $x$-position, so that the flux along the boundaries at $x_1$ and $x_2 $ is a factor $\phi$. 
 The vertical velocity at the lower boundary is assumed to be zero $v(0)= 0 $  
 and $ v(\phi) = \partial_t \phi $ at the upper boundary. 
 For an infinitesimal $\Delta x $ the volume integral can be replaced by a multiplication with the volume 
 $ \int_\Omega dV = \phi\Delta x $ . 
By this \eqref{movingVol} results in 
$ \phi \Delta x \partial_t \rho +  (\phi u \rho )_{x_2}  - (\phi u \rho )_{x_1}   +  \Delta x \rho \partial_t \phi = 0 $, 
or for $\Delta x \to 0 $ and with the help of the product rule 
\begin{align}
 \partial_t \phi \rho +  \partial_x (\phi u \rho )    = 0   
\end{align} 
In the same manner, the transport terms for momentum and energy produce the term necessary to include $\phi $ in the time derivative. 
Reconsidering all terms of (\ref{mass}-\ref{energy}) in this manner, one finds that only one further term is 
created by a time dependent $\phi$ which arises from the pressure work term in the energy equation. 
The same reasoning as above shows that the pressure work term 
$     
 \partial_x(   \phi  u {p} )  
 $ 
 becomes 
$ 
  \partial_x(   \phi  u {p} ) + p \partial_t\phi . 
  $ 
\begin{figure}[t]
 	\begin{center}
 		\includegraphics[viewport = 0 0 410 210, clip, width=.42\linewidth]{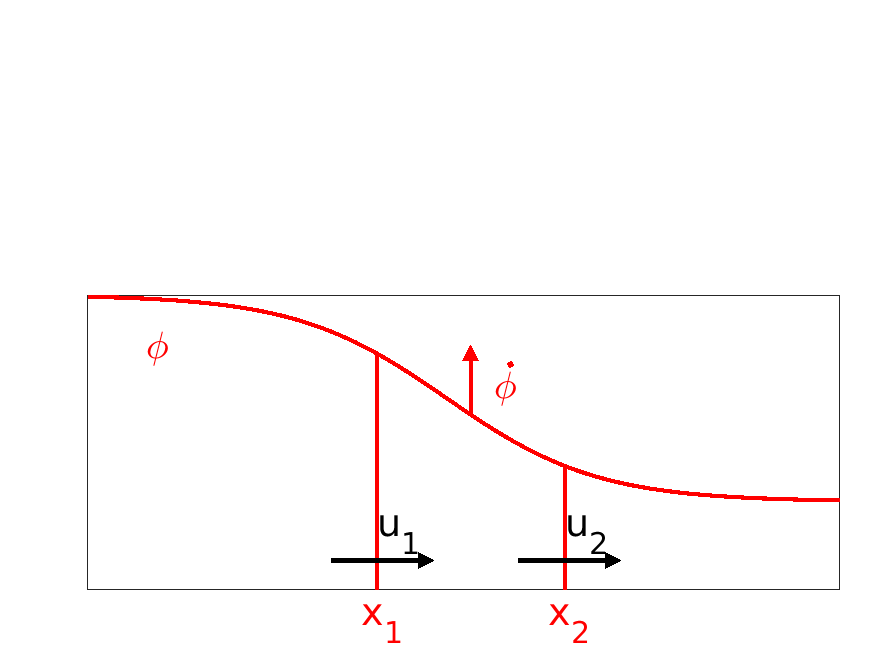}
 		
 	\end{center}	
 	\caption{\change{A time-dependent effective volume introduces addition flux terms, as $\dot \phi$ can be identified with a velocity at the boundary, see text. }}
 	\label{fig:effCrossDot} 	
 \end{figure}

The equations (\ref{mass}-\ref{energy}) become   
\begin{align}
	\partial_t (\phi \rho) + \partial_{x_\alpha}( \phi \rho u_{\alpha})   &=   0   \label{massMoving} \\ 
	\partial_t (\phi \rho u_\alpha) + \partial_{x_\beta}( \phi \rho u_\beta u_\alpha) + \phi \partial_{x_\alpha} p  &=  
	\phi  \chi  (u_\alpha^\mathrm{t} - u_\alpha  )
	+ \partial_{x_\beta} (\phi \tau_{\alpha \beta}) 
	 \label{momMoving}\\  
	\partial_t (\phi \rho e_t ) + \partial_{x_\alpha}[ \phi \rho u_\alpha ( e_t   +   u {p} ) ] + p \partial_t \phi&=  
	\partial_{x_\alpha} ( \phi u_\beta  \tau_{\alpha\beta} ) + \partial_{x_\alpha} (\phi \lambda \partial_{x_\alpha}  T)
	\label{energyMoving}
\end{align} 
A similar derivation for moving walls for the flow of blood is presented in \cite{AzerPeskin2007}, where the density becomes a constant due to the incompressibility assumption. 
The equations agree with Kemm et al. \cite{KemmGaburroTheinDumbser2020} if the Darcy and viscous terms are omitted. 

To elucidate the relevance of these extra terms, we use the product rule for time derivative terms, 
and splitting of the kinetic energy $  u_\alpha u_\alpha /2= e_t - e_\mathrm{in}  $ from the total energy equation to arrive at (in 1D)
\begin{align}
	\phi \partial_t ( \rho) + \partial_x( \phi \rho u)   &=   -  \rho \dot  \phi   \label{mass1DMovingSplit} \\ 
	\phi \partial_t ( \rho u) + \partial_x( \phi \rho u u) + \phi \partial_x  p  &=  -\rho u \dot \phi   \label{mom1DMovingSplit}\\  
	\phi \partial_t ( \rho e_\mathrm{in} ) + \partial_x( \phi \rho u  e_\mathrm{in}  ) + \partial_x( \phi  u {p} )  
	- \phi u  \partial_x p
	&=  -   ( p+ \rho e_\mathrm{in} ) \dot \phi .
	\label{energy1DMovingSplit}
\end{align} 
For the total energy of an ideal gas 
$e_t = (p/\rho)\cdot  1/(\gamma -1 ) $ the last equation becomes
\begin{align}
	\phi \partial_t ( p ) + \gamma \partial_x( \phi  u p   )  
	- (\gamma-1 ) \phi u \partial_x p 
		&=  -   ( \gamma p  ) \dot \phi 
\end{align} 
Without the spatial part a change in $\phi$ produces a change in the density given by 
$ d \phi /\phi + d \rho /\rho = 0 $ and $\phi  dp  + \gamma p d \phi  = 0 $ from which 
the adiabatic relation 
$d p/p - \gamma d \rho /\rho = 0 $ follows. 
This shows that a temporal changing $\phi$ is just an adiabatic pressure source, as expected for a homogeneous compression. 
Splitting off the mass equation from momentum shows that the change in momentum is fully due to the change of density. 
A detailed discussion of the dynamic $\phi$ is found in \cite{KemmGaburroTheinDumbser2020}. 
}

\paragraph{Implementation Strategy}
To describe embedded objects, the friction term $\chi $ can be chosen to be large or $\phi$ can be made small, where of course a combination is possible. 
The strategy will be to reduce the volume fraction by several orders of magnitude (typically down to $10^{-6}-10^{-8}$) so that the remaining volume is practically zero and by this all convective and viscous fluxes. Second, the Darcy friction is chosen as big as the time step permits for an explicit time marching. 
 \change{Stronger values the Darcy friction could be handled by an implicit or semi-implicit method as, for example, used in \cite{BoironChiavassaDonat2009}. 
 	However, a very large choice corresponds to simply setting the target values inside the objects. 
 	This reduces the smoothness of the solution having a detrimental effect on the discrete solution. 
 	This effect is discussed in more detail for example in \cite{EngelsKolomenskiySchneiderSesterhenn2015,vanyenKolomenskiySchneider2014}. 
 }
The reduced volume fraction implies a good approximation of the non-penetration condition, producing \change{thereby} a slip boundary condition, while the Darcy friction ensures the non-slip part. 
Close to a boundary, large shear forces result from this, restricting the time step. 
However, since the Darcy term needs to balance forces similar to the shear forces in the fluid itself, this leads to a very similar stiffness and no additional time step restriction results.

\subsection{Conservation properties} 

The conservation of mass and energy in the full domain is obvious from (\ref{mass}) and (\ref{energy}), while the momentum (\ref{mom}) is not conserved. 
However, in the case of immersed objects, it is not the conservation in the full domain $\Omega =  \Omega_\mathrm{f} \cup  \Omega_\mathrm{s}  $ that is physically relevant, but the conservation in the fluid $\Omega_\mathrm{f}$ domain without the objects $ \Omega_\mathrm{s} $. 
If we first consider a penalization by a sharply varying $\phi$ such that 
\begin{equation}
	\phi = \left\{ 
	\begin{tabular}{cl}
		$\epsilon$ &   $\Omega_\mathrm{s}$, inside the objects     \\
		1 &  $\Omega_\mathrm{f}$, otherwise 
	\end{tabular}\right. 
\end{equation}
and remark that the conserved quantities are weighted by $\phi$, we find e.g. for the mass
\begin{align}
	M = 
	\int_\Omega  \phi(\mathrm{x}) \rho(\mathrm{x}) dV 
	&= 
	\int_{\Omega_\mathrm{f} }  \phi(\mathrm{x}) \rho(\mathrm{x}) dV      
	+
	\int_{\Omega_\mathrm{s} }  \phi(\mathrm{x}) \rho(\mathrm{x}) dV\no\\
	&= % \approx 
	1 \int_{\Omega_\mathrm{f} }   \rho(\mathrm{x}) dV      
	+
	\epsilon \int_{\Omega_\mathrm{s} }   \rho(\mathrm{x}) dV  \label{decompCons}.
\end{align}
By this, the total mass is decomposed in mass in the fluid and in the solid region $ M = M_\mathrm{f} + \epsilon M_\mathrm{s} $. 
The last term is suppressed by a small volume fraction of $\eps$. 
If the value of the density in the porous region is bounded, its fraction of the total mass is suppressed by $\eps\to 0  $, 
the mass in the fluid domain is approximated by the total mass, and therefore approximately conserved.
In many practical cases, the maximal values of $\rho$ in the objects can be estimated from physical reasoning. 
This can be stagnation values in external flows or a doubling of pressure and density for acoustic reflections. 
We find very good conservation in tests, with the only ambiguity being the finite thickness of the transition of $\phi$ from
one to $\eps$, which is needed for numerical stability. 
This smearing of the boundary induces some arbitrariness of the exact location of the object and of the separation of the two integrals in \eqref{decompCons}.
However, mass leaving the fluid domain is not lost, but 'inside' the boundary region so that this property can be viewed as a not perfectly rigid boundary.   
The same reasoning can be made for the energy. The momentum is not conserved, as mentioned before, as objects act
as momentum sources by deflection flow or reflecting acoustics. 

By this, the volume fraction approach implies an (approximate) adiabatic boundary condition, if no temperature is forced.  
Since the energy content of the gas in the solid region is suppressed by $\phi$, its total energy content and the fluxes are suppressed by the same factor. By this, only negligible energy enters or leaves the object, assuming again finite values of the energy density in the object.

\subsection{Discrete equations} 
\change{Immersed boundaries allow using simple Cartesian grids.
	This permits to use finite difference methods, which are easy to implement and to parallelize. 
	Finite differences allow for small discretization errors by simply using appropriate finite difference stencils. 
	Various finite difference formulations are available.
	The here used method is a skew-symmetric method, due to the combination of strict conservation and the absence of numerical damping. 
	The first is essential for the faithful treatment of shocks, while the latter is advantageous for the simulation of acoustics and fine turbulence.  
	For skew-symmetric schemes, 
	the momentum transport term is formulated to yield a skew-symmetric matrix after discretization. 
	This has the elementary property that quadratic forms become zero $q^T A q = 0$ if the matrix $A$ is skew-symmetric $A=-A^T$. 
	This implies that the discrete momentum transport term does not change the kinetic energy, in agreement with the analytical theory. 
	A violation of this fact is a major source of instabilities since numerical damping or numerical excitation are equally likely. 
	
	It is found that the derivation of the skew-symmetric scheme of Reiss et al. \cite{ReissSesterhenn2014} can directly be adopted to include $\phi$. 
	Details can be found in the appendix~\ref{app:detail}, here we restrict the discussion to a brief description.
	However, this special discretization is not mandatory and it is expected that other discretizations work as well.

 }

We use $\sqrho, \sqrho u_\alpha$ and $p $ as calculation variables.
\change{This unusual choice origins from a symmetric splitting of the kinetic energy $ \rho u_\alpha u_\alpha/2= ( \sqrho u_\alpha)( \sqrho u_\alpha)/2   $. 
	This suggests using $\sqrho$ as a variable to form the velocity $ u_\alpha = (\sqrho u_\alpha)/\sqrho $, without evaluating the square root of a field. }
%We assume a time constant volume fraction $\phi$. 
\change{The final equations (for a time constant volume fraction $\phi$) are }% final result is 
\begin{align}
	\partial_t ( \sqrho) + \partial_{x_\alpha}( \phi \rho u_\alpha)/(2\phi\sqrho)    &=   0   
	\label{massSkew} \\ 
	\phi \sqrho	\partial_t ( \sqrho u_\alpha) + 
	\frac 1 2 \left[ \phi \rho u_\beta \partial_{x_\beta}  u_\alpha +  \partial_{x_\beta}( \phi \rho u_\beta u_\alpha)\right)]   + \phi \partial_{x_\alpha} p  &= \no\\
	\phi  \chi  (u_\alpha^\mathrm{t} - u_\alpha  )  
	+ \partial_{x_\beta} (\phi \tau_{\alpha \beta})  &
	\label{momSkew}\\  
	\frac{\phi}{\gamma -1 } 	\partial_t ( p ) + \frac{\gamma}{\gamma -1 }\partial_{x_\alpha}( \phi p u_\alpha    ) - \phi u_\alpha \partial_{x_\alpha} p &=  \no\\
	\partial_{x_\alpha} ( \phi u_\beta  \tau_{\alpha\beta} ) + \partial_{x_\alpha} \phi \lambda \partial_{x_\alpha}  T 
	- u_\alpha \left[   \phi  \chi  (u_\alpha^\mathrm{t} - u_\alpha  )   	+ \partial_{x_\beta} (\phi \tau_{\alpha \beta})   \right]&.
	\label{energySkew}
\end{align} 
\change{The equations are discretized simply by replacing all derivatives with finite differences with symmetric stencils and evaluating all products or divisions pointwise.
	The time-stepping is the standard fourth-order Runge-Kutta scheme. 
	This is followed by filtering for cases where otherwise highly oscillatory solutions occur, 
	namely when complex geometries focus waves and most importantly for shocks where 
	dissipation is demanded by physics.
	A specially designed filter is needed in general, which is discussed in section~\ref{sec:filter}. 
	
	The whole method is fully explicit and simple to implement. 
	%A source code producing all the numerical examples is provided as supplementary information on the journal 
	%homepage %\footnote{\color{red} Remark for the reviewers: The code will be put online as \emph{Supplementary Information}
	%	for the final version of the publication to match the final version. }. 
  }

 It seems attractive to include 
$\phi$ in redefining $\rho$ and $p$, but big rounding errors were found in this case for a small $\phi$, so we refrain from doing so. 
Note that we divide by $\phi$ for time-stepping, so that it cannot be set to zero in the numerical implementation.  
The transport term in the momentum equation is explicitly skew-symmetric, which is preserved if the spatial discretization simply replaces the derivatives with central finite difference matrices. 
The kinetic energy is changed only by compression work, and not by numerical friction. The total energy as the sum of the kinetic and internal energy is conserved. 

The skew-symmetric form suggests an unconditionally stable scheme. 
For this, a time-stepping that conserves quadratic invariants is needed as presented by Brouwer et al. \cite{BrouwerReissSesterhenn2014b}. 
To be rigorously norm-stable, the variable $\sqrt{p}$ instead of the pressure $p$ has to be used, however, this form works very well in our experience. 
All methods which conserve quadratic invariants are implicit methods. 
This extra effort does not seem to be justified for many time dynamic simulations, where all scales of interest need to be resolved since stability does not guarantee the correctness of the fast dynamic. 
Explicit high order time-stepping yields very good conservation properties, so that it can be used in most practical cases \change{as shown for different examples in \cite{BrouwerReissSesterhenn2014b}. 
	The conservation can cheaply be validated during simulation permitting to switch to a different time integrator. 
	A reduction of the time step might often be sufficient as thereby the conversation is strongly improved \cite{BrouwerReissSesterhenn2014b}.
	Explicit time integrators with improved energy convergence are discussed for incompressible flows by Capuano et al. \cite{CapuanoCoppolaRandezLuca2017}, which might be adopted for our method. 
 }

\section{Filtering}

\label{sec:filter}
The non-linearity of the Navier-Stokes equations constantly produces higher wavenumbers, which tend to exceed the grid resolution after some time if not removed by friction terms.
This is especially important in shocks, but also at boundaries. 
The high gradient of the immersed objects can create high-frequency structures up to a grid-to-grid oscillation. 
\change{To remove these oscillations, artificial dissipation is often applied.
	Various methods to introduce dissipation exist, an overview is given by Pirozzoli \cite{Pirozzoli2011a}. 
	In the simplest case it is introduced by upwind discretizations or similar by calculating fluxes by a Riemann solver, which creates in effect an upwind stencil for the Riemann invariants \cite{LeVeque1992}. 
	
	Often hybrid methods are used, which switch based on a flux limiter from a high order, low dissipative to low order dissipative scheme. 
	Alternatively, artificial viscosity by an adaption of the viscous parameters near discontinuities was proposed by Cook et al. \cite{CookCabot2005}. 
	Artificial viscosity and the flux limiting are at least for simple examples structurally similar, see \cite[Chap.~16]{LeVeque1992}. 
	Adaptive filtering was introduced with the same purpose by Yee et al. \cite{YeeSandhamDjomehri1999}, where a filter operation with local filter strength allows to concentrate dissipation near discontinuities. 
	With a similar effect, 
	%, or with a similar effect with
	varying filter order can be applied as used by Visbal et al. \cite{VisbalGaitonde2005} and more recently by Patel et al. \cite{PatelMathew2019}. 
	A quantitative comparison of the different methods is provided by Johnsen et al. 
	\cite{JohnsenLarssonBhagatwalaCabotMoinOlsonRawatShankarSjoegreenYeeZhongLele2010}. 
	These non-linear dissipation methods rely usually on a detector of shocks or non-smoothness, which is used to determine the flux limiter, the artificial viscosity, or the filter strength. 
	Different shock detectors are compared in \cite{Pirozzoli2011a}.

\bigskip 

%\subsection*{The adaptive and conservative filter }
 }

In the following, a filter is used in the numerical examples.
This should not only remove oscillatory components but also allow a locally varying filter strength and especially respect the conservation with the varying volume fraction. 
Conservative filtering is often achieved by filtering the conserved quantities, however, this is not appropriate here. 
Consider the mass in one dimension, $M = \sum_{i } (\rho \phi)_i \Delta x$, where $i$ runs over all grid points. Filtering $(\rho \phi)_i$ would keep the total mass constant, however, it would tend to increase $\rho$ where $\phi$ is becoming small, i.e. inside the boundary of immersed objects. This counteracts the intention to smooth the physical \change{quantities}. 

Conservative filters similar to \cite{KimLee2001} are now formulated analogously to dissipative flux terms, which fulfill the desired properties, as argued in the following. 
The starting point is to define fluxes between two adjacent grid points, similar to finite volume for some variable $q$ (depicted in 1D)
\begin{align}
	\overline { (\phi_i q_i )} = (\phi_i q_i) +   (F_{i-1/2} - F_{i+1/2}    )       \label{dispFilter}, 
\end{align} 
with $F_{i-1/2} = -(\sigma\phi)_{i-1/2} ( q_i - q_{i-1}  )/4    $ and $F_{i+1/2}  = F_{(i+1)-1/2}  $. 
The flux form guarantees the conservation of $\phi q$ in the filtering, while the flux is non-zero only if $q$ varies and the 
local flux factor $\sigma_i$, $(\sigma\phi)_{i-1/2}  =(\sigma_{i-1}\phi_{i-1}+ \sigma_{i}\phi_{i})/2   $ is non-zero. 
The filter strength $\sigma_i$ can be locally varying and is thereby chosen to filter oscillations within objects or in under-resolved parts of the flow, e.g. close to shocks. 

To generalize the dissipation-like filtering (\ref{dispFilter}), it can be written in matrix notation as
\begin{align}
	\overline {  (q\phi  ) } =  (q\phi)  +   \bar D  (M (\sigma\phi) ) D q    \, ,  
\end{align} 
where the expression (\ref{dispFilter}) is recovered with 
\change{$ (\bar D u )_i =   (u_{i} - u_{i-1})/2  $ and $ ( D u )_i =  (u_{i+1} - u_{i})/2 $} and the averaging operation $ (M u )_i = (u_{i-1} + u_{i})/2  $. 
\change{For the derivation periodic grids are assumed from which the non-periodic case is constructed at the end.}
This method is second-order if $\sigma$ and $\phi$ are constant, and first order otherwise. 
This form reveals the principle structure whereby the damping and the conservation properties can be checked easily, as now detailed. 
With $\bar D  = -  D^T  $, and a positive semi-definite matrix $(M(\phi \sigma))$ (in our case a diagonal matrix with non-negative elements), we find that the operator $\bar D  (M(\phi \sigma)) D$ is negative semi-definite, since 
\begin{align}
	q^T \bar D  (M(\phi \sigma)) D  q = - (D q)^T (M(\phi \sigma) (D q)  \leq 0  .
\end{align}

Obviously $\overline {  (q\phi  ) } =  (q\phi)$ holds, if $q$ is constant, since the derivative vanishes for any constant function $ D {\bf 1 }  =0   $, with the vector of constant entries $({\bf 1})_i = 1 $, used to depict the sum. 
The conservation is checked by 
\begin{align}
	{\bf 1}^T \overline {  (q\phi  ) } =  {\bf 1}^T(q\phi)  +  {\bf 1}^T \bar D  (M (\sigma\phi) ) D q     =  {\bf 1}^T(q\phi)  
\end{align} 
due to the telescoping property $ {\bf 1}^T \bar D =0 $ \change{for periodic grids}. 

To generalize \change{ to higher order}, appropriate $\bar D $ (thereby $D=-\bar D^T$) and $M$, fulfilling the mentioned properties, can be chosen. %the method can be generalized. 
We demand that the resulting filter yields the filter stencil of a higher-order filter when $\phi$ and $\sigma$ are constant. 
In \cite{BogeyCacquerayBailly2009} $\bar D$ was kept proportional to a one-sided first derivative to keep the flux form, and $D$ was calculated to create the desired filter stencil. 
Here, we instead chose to keep $\bar D  = -  D^T  $ so that the negative semi-definiteness is guaranteed. 
We remark that the stencil $(1, -2, 1 )/4  $ for $D $ and $\bar  D = -D^T$ yields the filter $   (-1,  4, -6, 4,  -1)/16 $ for constant $\phi\sigma$, which is the standard fourth order filter, reducing to second-order for non-constant $\phi$ or $\sigma$. 
It has the nice property %The obvious advantage, in contrast to the work \cite{BogeyCacquerayBailly2009}, is
that a purely linear function is reproduced by filtering, even for non-constant $\sigma$ or $\phi$.  
Similarly, using this filter stencil for $D$ we find the filter stencil of eighth order 
$( -1, 8,  -28,  56,  -70,  56,   -28,   8 ,      -1)/256$ for constant $\sigma$ and $\phi$. 
In both cases, no averaging by $M$ is done, i.e. $M =1 $. 
The averaging in the first/second-order method is needed to avoid a directional diffusion if the stencil $D$ is non-symmetric. In more dimensions, the contributions of the different spatial directions are simply added
\change{
	where different $\sigma_\alpha$ for all spatial directions $\alpha$ are possible. 
	An algorithmic description and the later used adaptive filter strength for shocks are discussed in the appendix~\ref{app:filter}.} 

\change{The non-periodic case can be handled with the same $D$ and $M$ as above
	by choosing  $\sigma=0$ close to the boundary for a sufficient number of grid points, which results in a non-periodicity stencil of the resulting filter. 
For the first/second-order filter, the choice of $\sigma=0$ for the two points next to the boundary in the direction perpendicular\footnote{In calculation space for transformed grids.} to the boundary does not change the boundary values. 
%Additionally $\sigma$ for the filtering components 
Setting only
the outermost points of $\sigma$ zero would result in modified outermost points, but without a modification of the flux by the filter \cite{HossbachLemkeReiss2021}.
This can be chosen depending on the desired boundary condition.
 } 

%\newpage

\section{Numerical examples}
\label{sec:examples}
\subsection{Acoustic Configurations} 
\label{sec:ac}
\subsubsection{Acoustic Reflection in 1D}
\label{sec:ac1Dpulse}
\begin{figure}[ht]
	a) ~~~~~~~~~~~~~~~~~~~~~~~~~~~~~~~~~~~~~~~~~~~~~~~ b)\\ 
	\includegraphics[width=.48\linewidth]{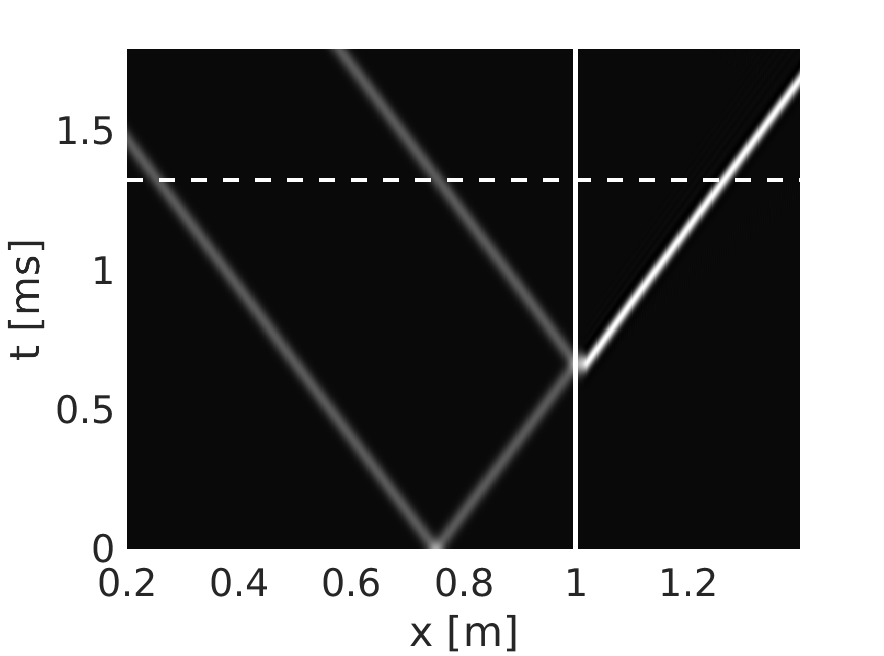}
	\includegraphics[width=.48\linewidth]{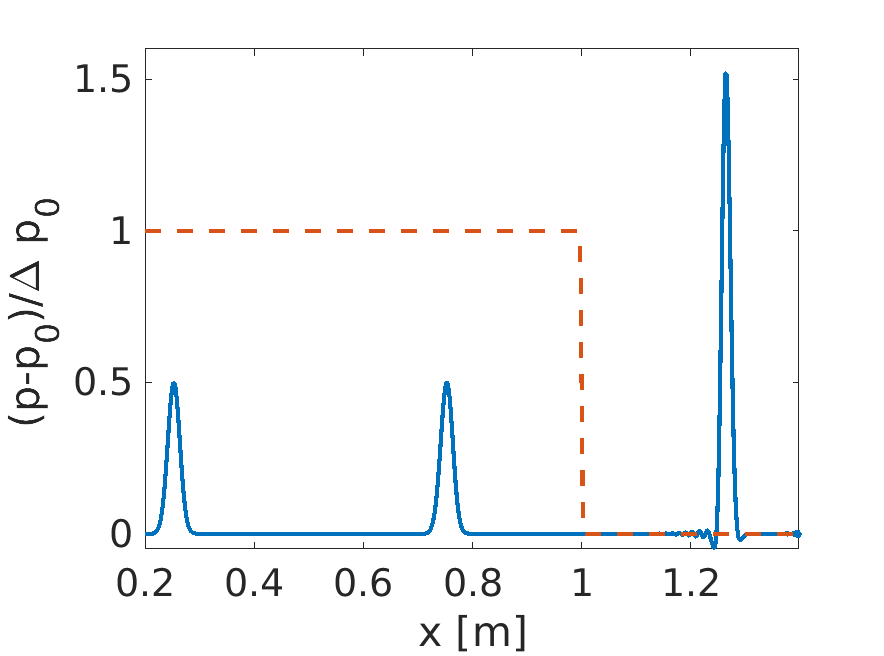}
	\\	
	c) ~~~~~~~~~~~~~~~~~~~~~~~~~~~~~~~~~~~~~~~~~~~~~~~ d)\\ 
	\includegraphics[width=.48\linewidth]{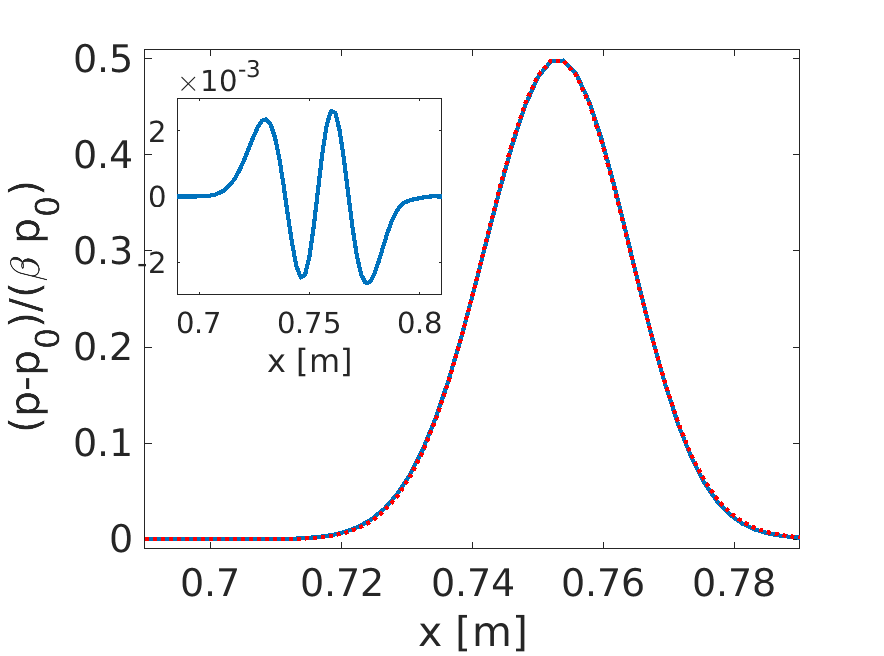}
	\includegraphics[width=.48\linewidth]{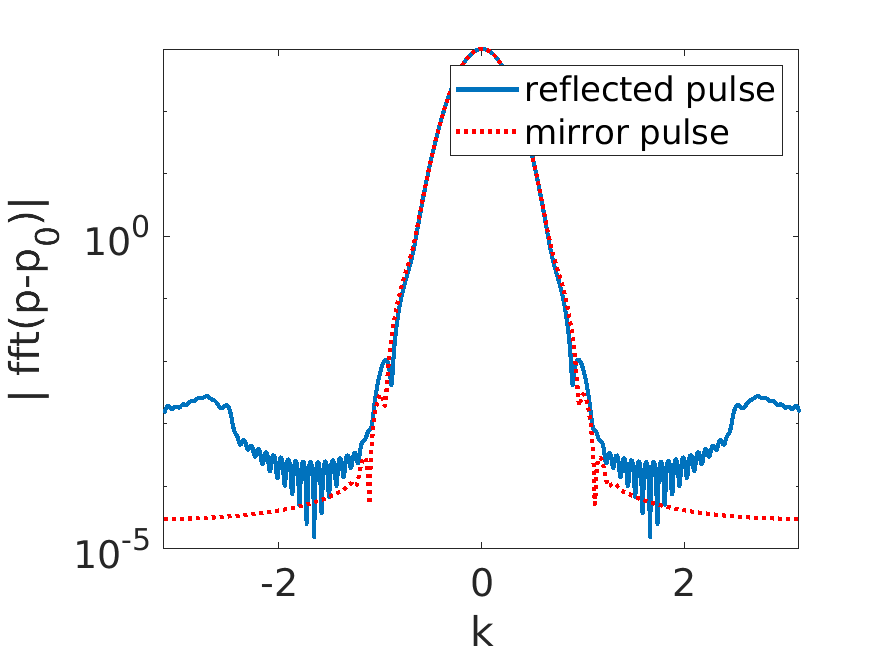}
	
	\caption{a) Space-time diagram of an acoustic pulse ($p$). Immersed boundary depicted by a dotted line, time of snapshot by broken line. 
		b) Snapshot of pressure scaled by initial pulse strength, volume fraction dotted. 
		c) Reflected pulse in comparison with mirror pulse. Relative difference in the inset. 
		d) Spectrum of pulse shows good agreement over a large part of the wavenumbers. 
	}
	\label{fig_1dpulse}
	% immersedObjects/acoustic/adiabaticPulse1D/03	
\end{figure}

A solid boundary reflects acoustic pulses. 
Here, we investigate how such a reflection is created by the volume fraction alone. % of an acoustic pulse. 
The one-dimensional domain with a length of $L=2$m is described with $N=1024$ points. 
The domain is large enough to ignore the domain boundaries for this example. 
The wall starts at $x_0=1\mathrm{m} + \delta_0$ where only the volume fraction is reduced from 1 to $\epsilon = 10^{-8}$ with a hyperbolic tangent 
\begin{align}	
	f(x) = 1- (1-\epsilon) (1+ \tanh( (x-x_0)/\delta ))/2   
	\label{boundFun}
\end{align} 
with a thickness of $\delta  = \Delta x$ as the grid spacing. 
The shift $\delta_0 = 0.3 \Delta x$ is chosen to align the maximum of the reflected and the mirror pulse by visual inspection.
This implies that the effective location of the wall is not at \change{$\phi \approx 1/2 $} but shifted for the fraction of a grid spacing. 
No Darcy term or filter is used in this example to reveal the effect of the volume fraction. 
Standard fourth-order central difference and the classical Runge-Kutta-4 time stepping is applied, 
\change{as detailed in the appendix~\ref{app:detail}}. 

The initial data is an adiabatic, Gaussian pulse in pressure and density,\change{
\begin{align}
\rho = \rho_0 \left(1+ \beta\exp\left(- (X-x_0)^2 /\sigma_\mathrm{pulse}^2\right) \right)
\qquad 
 p =  p_0  (\rho/\rho_0)^\gamma 
\end{align}
 with relative height of $\beta= 10^{-3}$,} relative to $\rho_0=1$kg/m$^3$, $p_0= 10^{5}$Pa and $\sigma_\mathrm{pulse} = 8 \Delta x$, 
which is chosen so that no artefacts of the numerical dispersion error are visible by inspection. 
For $\sigma_\mathrm{pulse} = 6 \Delta x$, the pulse is followed by clearly visible ringing.

The results are presented in Fig.~\ref{fig_1dpulse}. 
The space-time diagram shows that the right-going pulse is reflected at $x=1$m. 
The speed of sound is clearly the same within the object, as well as outside. 
The snapshot of $p$ at $t\approx 1.32$ms is shown in the second plot. 
The left-going and the reflected pulse have the same amplitude, as expected, while the pulse in the object is strongly increased and shows some ringing. Its increase is due to the doubling of the pressure at the reflection point. 
Some additional increase is created by the restriction to very small $\phi$, which is not further analyzed here. 
To quantify the quality of the reflection, the reflected pulse is compared with a mirror pulse, and a very good agreement is found. The inset shows the difference relative to the initial pulse amplitude $\sim 2 \cdot 10^{-3}$. 

\change{  
	Finally, the spectrum is shown to reveal the frequency dependency of the difference between the reflected pulse and the reference mirror pulse.
	A window $[x_m - 5.5 \sigma_\mathrm{pulse} ,x_m +5.5 \sigma_\mathrm{pulse}   ]  $ is created around location the maximum of reflected pulse $x_m$. 
	The pressure values outside this sharp window are set to zero, and a Fourier transform is calculated for the remaining values.
	By this, the spectrum of the reflected pulse and the reference pulse is calculated. 
	In Fig.~\ref{fig_1dpulse} the absolute value of these two is compared. 
	The numerically interesting range is very well represented. 
	Since the absolute value agrees very well, the error visible in Fig.~\ref{fig_1dpulse}c is dominantly a phase error. 
} 

All results are largely independent of $\eps$, if sufficiently small. It can be increased by orders of magnitude without changing the result or decreased further without compromising the numerical stability. The only visible effect is that the ringing of the pulse inside the object and the kink at the boundary becomes stronger with a smaller $\eps$. 

The observed displacement $\delta_0$ is much smaller in comparison with the in-depth discussion of \cite{HesterVasilBurns2020}. 
However, a substantially different situation was considered there: an incompressible boundary layer enforced by a pure Darcy term. 

\subsubsection{Containment}

\begin{figure}[ht]
	\begin{center}
		\includegraphics[width=.55\linewidth]{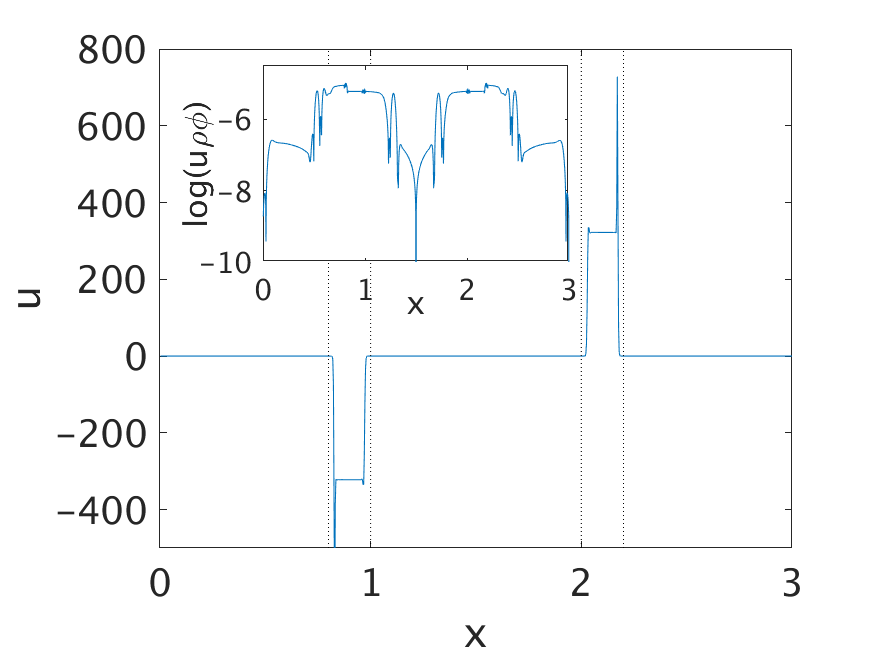}	
	\end{center}
	\caption{ A one dimensional domain with two walls at $x=1$ and $x=2$ holding a fourfold pressure compared with the outside. The start and end of the walls is marked by the thin broken lines. 
		The volume fraction allows to contain the gas: while the flow accelerates to the speed of sound (chocked flow), the mass flow 
		$\rho u \phi $ is suppressed by the volume fraction $\phi =10^{-8}$. A Darcy term can reduce the flow further. The inset shows the logarithm of the mass flow, where small residual acoustic noise is visible. } 
	\label{fig:tightness} 
\end{figure}

One motivating aspect of this investigation is to improve the penalization for configurations with high-pressure gradients. 
If the Darcy term is the only term to combat a high-pressure difference, it must be chosen accordingly large, leading to a strong numerical stiffness or even instabilities.
In the method proposed here, the pressure gradient is counteracted by a change in volume fraction, 
which is largely able to handle the pressure gradient even without a Darcy term. %fig~\ref{fig:effCross}.
In Fig.~\ref{fig:tightness}, a set-up is shown where two walls at $ x=1$ and $x=2$ keep a high pressure reservoir of 4$p_0$ against $p_0 = 10^5$. 
Without a Darcy term, a chocked flow is found. 
The flow accelerates to sonic conditions of Ma $= u/c \approx 1 $, with the sound velocity $c=\sqrt{\gamma p /\rho} $, creating a supersonic expansion terminated by a shock. 
Chocking conditions are not fully reached in the wall (Ma $\approx0.75$), which is due to the filter acting similar to a viscosity. 
The second/fourth-order filter described in Sec.~\ref{sec:filter} is applied in the whole region. 
However, since the volume fraction $\phi = 10^{-8}$ scales the mass flux, it results in a mass flux of the order of $\approx 10^{5}$. 
This is largely independent of wall thickness. 
The steady-state is not fully reached in the simulation, instead, small acoustic noise is still visible in the mass flux in the inset of Fig.~\ref{fig:tightness}.
A Darcy friction term allows reducing the mass flux further. 
The findings show the ability of the method to handle large pressure differences.

\subsubsection{Stiffness of the method} 
\label{sec:stiff} 
Based on the CFL number criterion, it was argued that leaving the speed of sound unaltered implies avoiding additional stiffness \cite{KevlahanDubosAechtner2015}. 
However, since the pressure amplitude is rapidly increased and the velocity is reduced when a sound pulse reaches the wall location, some additional stiffness is expected. 
The magnitude of the source term in the momentum equation (\ref{mom1D}) suggests the possibility of a massive increase. 
Yet, a very moderate decrease of the admissible CFL number, depending on the width of the boundary function, was found in the simulation. 

\begin{figure}
	\begin{center}
		\includegraphics[width=.55\linewidth]{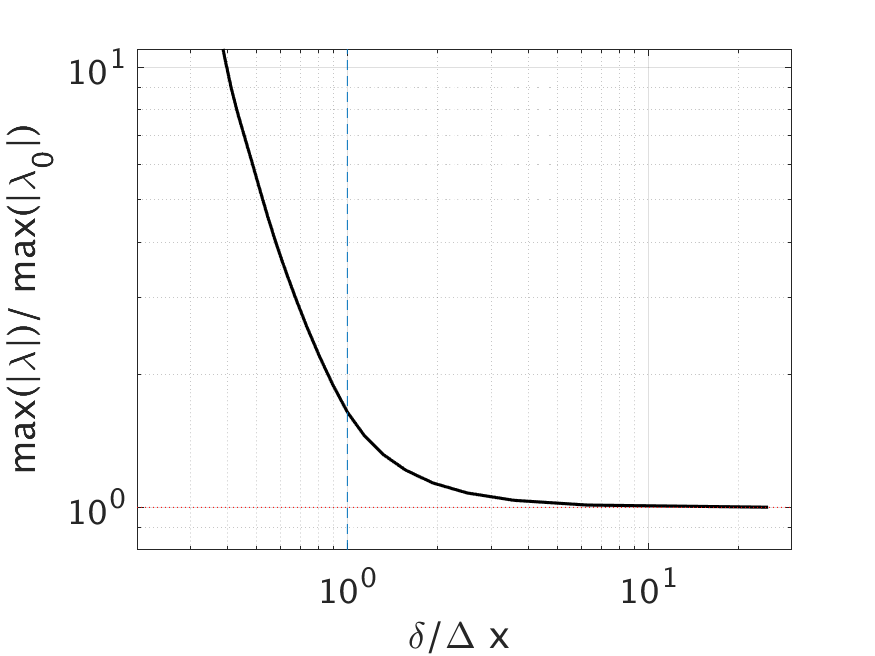}	
	\end{center}
	\caption{The increase in stiffness is given by the increase of the maximal eigenvalue modulus as a function of the boundary smoothing thickness $\delta$. 
		It is normalized by the maximal eigenvalue of the problem with constant $\phi$. For wide boundary functions ($\delta\gg \Delta x$) the stiffness agrees with the unmodified system ($\phi\equiv1$), while for 
		small values of $\delta$ a steep increase is visible.  } 
	\label{fig:stiff} 
\end{figure}

To investigate the stiffness, the eigenvalues with maximal magnitude were determined for the described acoustic setup of section \ref{sec:ac1Dpulse}, see Fig.\ \ref{fig:stiff}. 
For a wide wall function $\delta\gg \Delta x$ we reproduce the stiffness of the right-hand side without the volume fraction $\phi\equiv1$. For small $\delta$, a sharp increase of the stiffness was found.
For the setting of the simulation $\delta = \Delta x $ an increase of roughly $70\%  $ was found, consistent with the possible CFL numbers in the simulation. 
For a slightly stronger smeared boundary $\delta = 1.5 \Delta x $, an increase of 25\% of the stiffness was found. As expected, the eigenmodes with the largest eigenvalues are always strongly localized at the boundary location. 
One conclusion from this is that in order to increase the quality of the boundary, instead of decreasing $\delta$, a finer grid resolution can be less restrictive on the time step. 
Altogether, the increase is moderate and can often be much smaller than a pure Darcy term, especially for high-pressure gradients. 

\subsubsection{Acoustic Reflection in 2D}

\begin{figure}
	a) ~~~~~~~~~~~~~~~~~~~~~~~~~~~~~~~~~~~~~~~~~~~~~~~ b)\\ 
	\includegraphics[width=.48\linewidth]{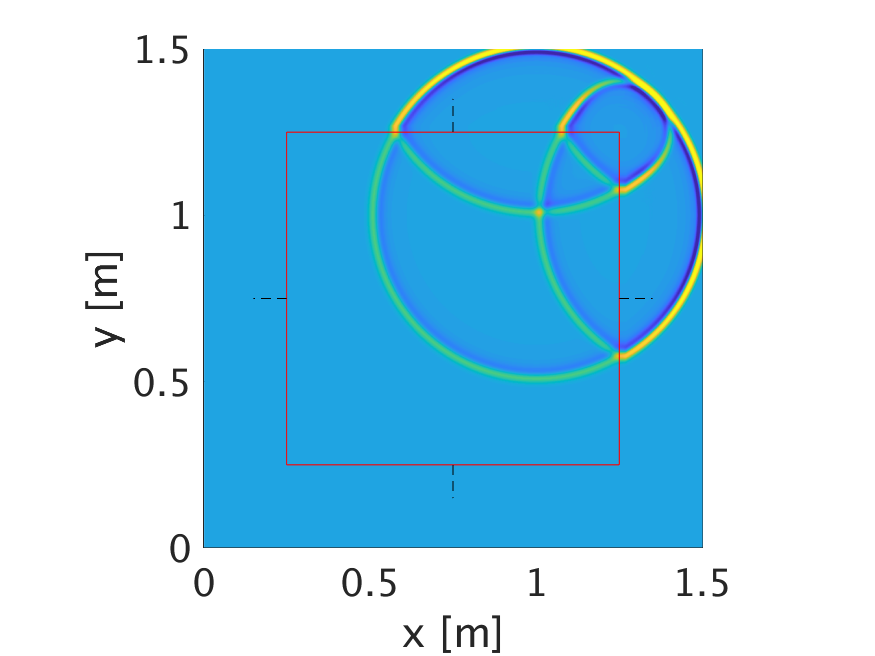}
	\includegraphics[width=.48\linewidth]{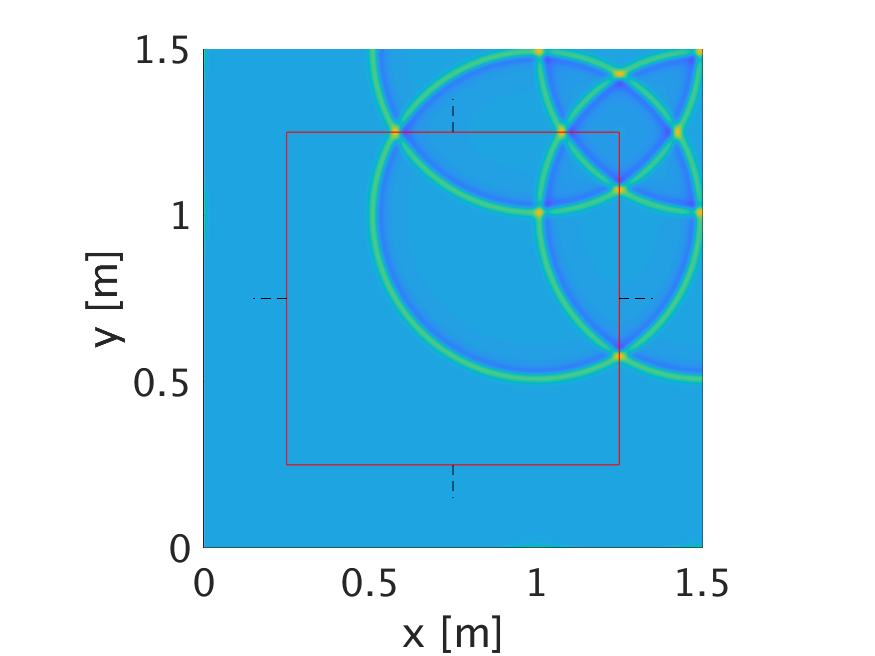}
	\\	
	c) ~~~~~~~~~~~~~~~~~~~~~~~~~~~~~~~~~~~~~~~~~~~~~~~ d)\\ 
	\includegraphics[width=.48\linewidth]{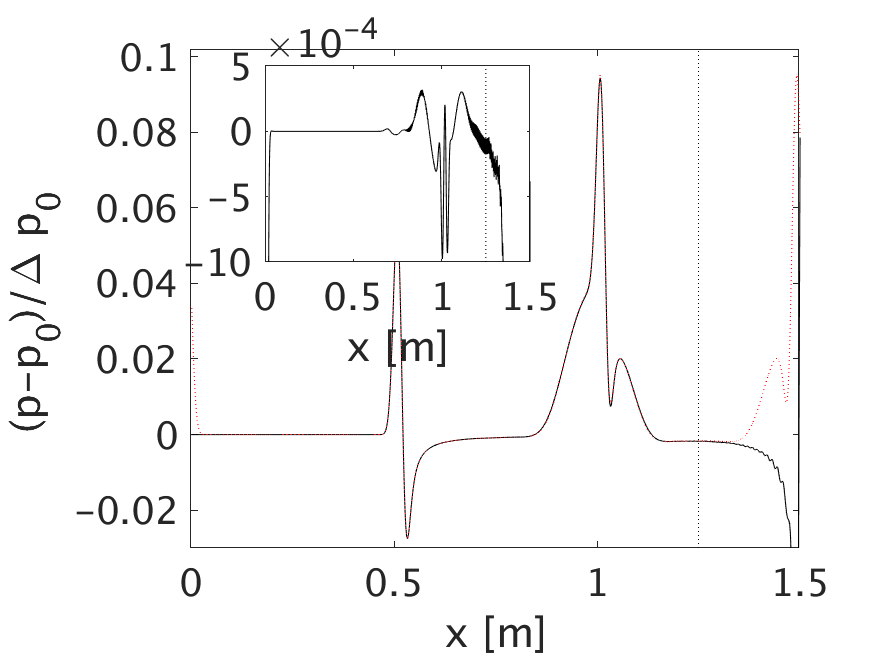}
	\includegraphics[width=.48\linewidth]{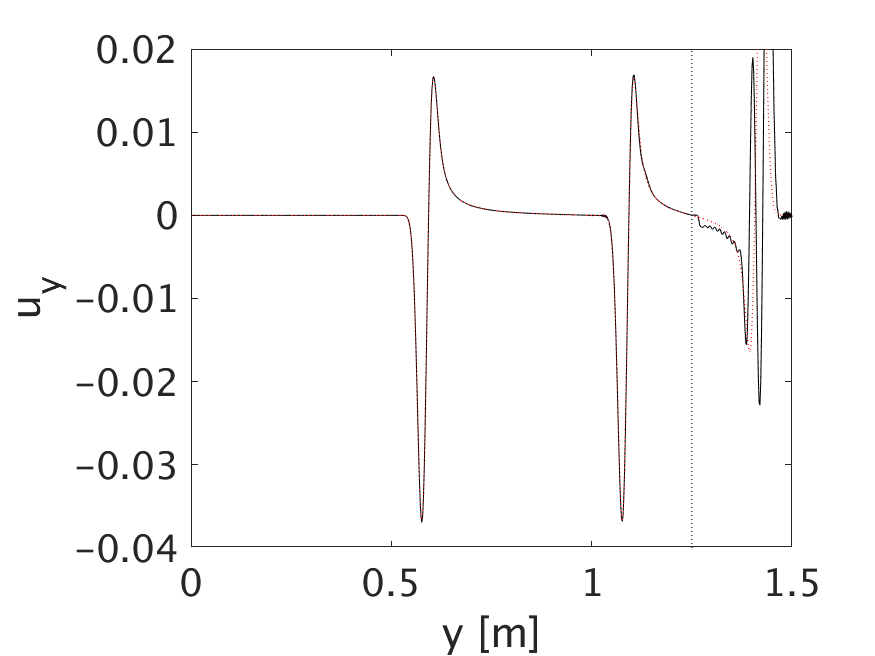}
	
	\caption{a) Snapshot of the pressure, the boundary depicted by a thin red line. 
		b) Mirror pulse solution. 
		c) A cut along $y=1$, to allow a comparison, difference in the inset. 
		d) The tangential velocity along $x=1.25$ agrees excellently with the reference showing a faithful construction of slip boundaries. 
	}
	\label{fig:2dpulse}
\end{figure}

The acoustic case is extended to two dimensions. A square domain $x\times y  =(1/4,\dots,5/4)\times (1/4,\dots,5/4)$ is embedded
in a domain of size $2~\mathrm{m}\times2~\mathrm{m}$, discretized with $1024\times 1024 $ points.
An adiabatic pulse centered at $(1~\mathrm{m},1~\mathrm{m})$, again with $\sigma_\mathrm{pulse} = 8 \Delta x$ and an amplitude of $10^{-3} \rho_0$, $\rho_0=1$, is used as initial condition. 
A reference can again be created by the three mirror pulses, \change{located at (1.5 {m}, 1 {m}), (1 {m},1.5 {m}), (1.5 {m}, 1.5 {m}) with the same amplitude as pulse inside the box.
	The pulse is reflected by the jump in $\phi$, modeling  the wall. 
	Again, acoustic waves are created within the object, where for a locally constant $\phi=10^{-8} $ in effect the same Euler equations are valid. 
	The mass, momentum, and energy content of the wall region are small by the factor of $10^{-8}$ as discussed before.} 
Again, we find excellent agreement, see Fig.\ \ref{fig:2dpulse}. 
Not only is the acoustic wave, which impinges perpendicular to the wall, well reproduced, furthermore the wave slipping along the wall at $x= 1.25$ is correctly reproduced. Such a slip boundary cannot be presented if a Darcy term is used to describe objects since this necessarily creates non-slip boundaries.

\subsection{Flow configurations}
\label{sec:flows}
\subsubsection{Potential Cylinder Flow} 
\label{sec:potFlow}
To investigate the quality of the slip boundary condition, the friction-free flow around a cylinder is calculated, for which exists a simple analytical solution. 
A flow velocity at infinity of $u_0= 10$ m/s is used, so that (for reference values as above) a low Mach number of Ma $\approx 0.02$ allows to consider the flow as near incompressible. 
As boundary conditions, the known potential flow 
$u_\mathrm{pot}= \partial_x  \Phi_\mathrm{pot} $ and 
$v_\mathrm{pot}= \partial_y \Phi_\mathrm{pot} $, with the potential in polar coordinates relative to the cylinder center $\Phi_\mathrm{pot} = u_0 \left(1 + \frac{r_0^2}{r^2} \right ) \cos(\vartheta)   $, is set. 
The pressure is calculated from the Bernoulli theorem, the density by the adiabatic condition.

\begin{figure}	
	\includegraphics[width=.6\linewidth,clip = true, viewport= 10 50 400 250]{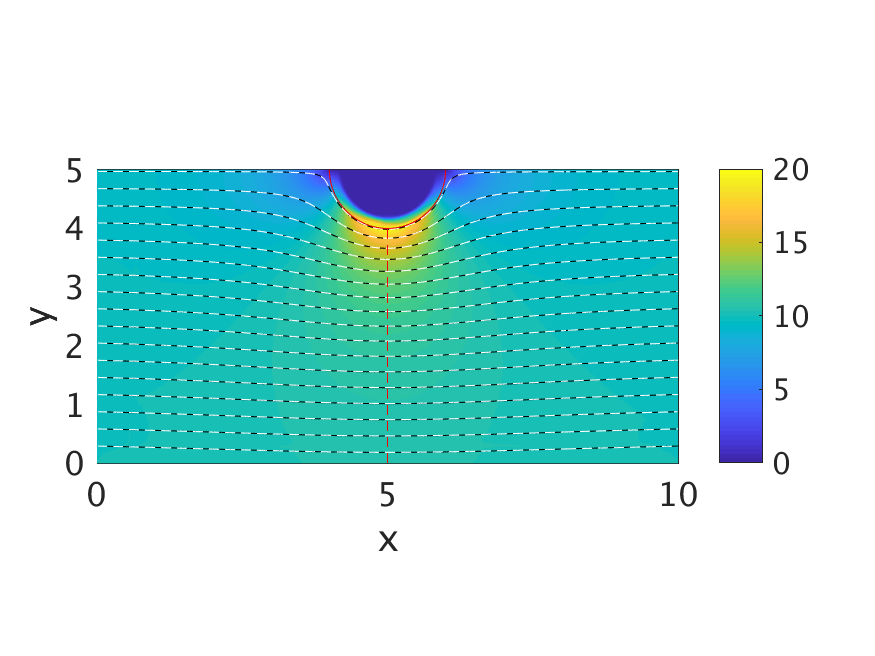}
	\includegraphics[width=.38\linewidth,clip = true, viewport= 10 0 400 300 ]{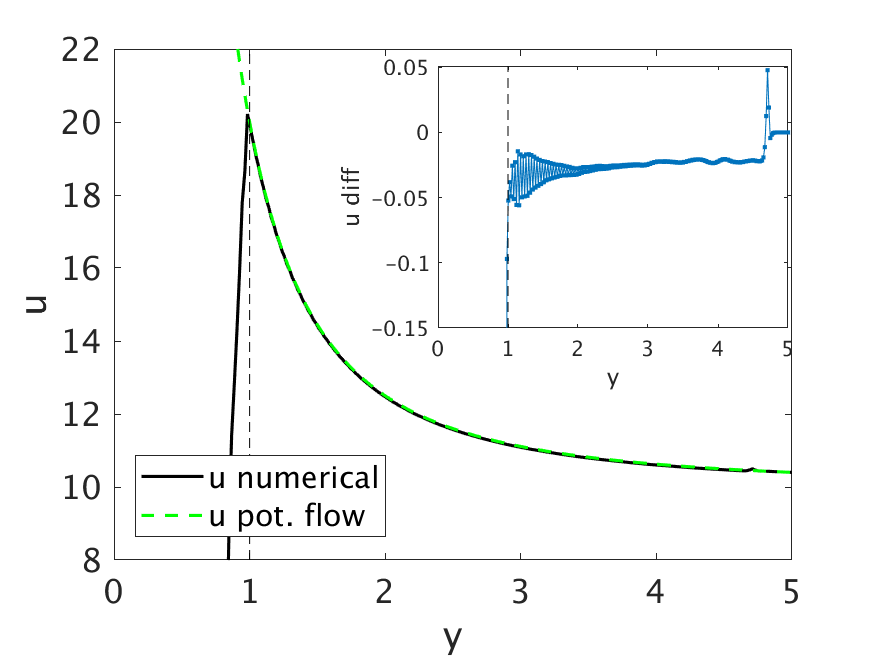}
	\caption{A flow around a slip cylinder compared with a potential flow solution.
		Left: The numerical streamlines (broken, white) are almost perfectly on top of the analytical reference, depicted in black, the velocity magnitude is color-coded. 
		Right: The $x$ component of the velocity at the line ($x=5$, broken red line in left plot) in comparison with the analytical value, shows very good agreement. The difference shows kink close to the outer sponge ($y\approx 4.4$), where the analytical solution is enforced and a zigzag mode. }\label{potentialCylinder}
	%/USER/reiss/immersedObjects/cylinder/07
\end{figure}

The 2D domain of size $(10~\mathrm{m}\times 10~\mathrm{m})$ is discretized by $514\times 514$ points and contains a cylinder of radius \change{$R_0=1~\mathrm{m}$}. 
The wall of the cylinder is again smoothed by the same hyperbolic tangent as above with $\delta = 1.5\Delta x$. 

The initial condition is the velocity constant $(u_0,0) $ (impulsively started cylinder). 
At the boundary, we force the analytical solution by a sponge layer of thickness $10\Delta x$, 
and use a forcing strength for the Darcy and the sponge term of $\eta = 2 \Delta t$.  

Further, a filter of first/second-order as described above is used only inside the cylinder, again with the same mask function retreated by $4\Delta x $ inside the object. 
A Darcy term is used with a mask created by the same function as the volume fraction (\ref{boundFun}), where the reference point $\bar x_0$ is moved relative to the wall position by $7\Delta x$ so that its action is practically zero at the boundary.
The purpose of both is to avoid dynamics inside the object, which otherwise destabilizes the simulation. 

We find excellent agreement with the analytical solution, see Fig.~\ref{potentialCylinder}. 
The numerical streamlines are on top of the analytical ones and the pattern of the velocity magnitude matches the expectations. 
The $x$ component of the velocity along $x=5$ also shows a very good quantitative agreement. A kink at the position where the sponge enforcing the boundaries ends, and a small zigzag mode relative to the analytical solution, is visible.

The differences might be explained by a slightly different blockage, due to the non-sharp boundary location. 
This interpretation is supported by the fact that the error becomes smaller with higher grid resolution, 
by which this effect becomes effectively smaller. It is concluded that slip conditions are created which are fully sufficient for practical calculations.

\subsubsection{Wedge in supersonic flow}

\begin{figure}[t]
	
	\includegraphics[width=.49\linewidth,viewport = 70 0 340 300,clip = true]{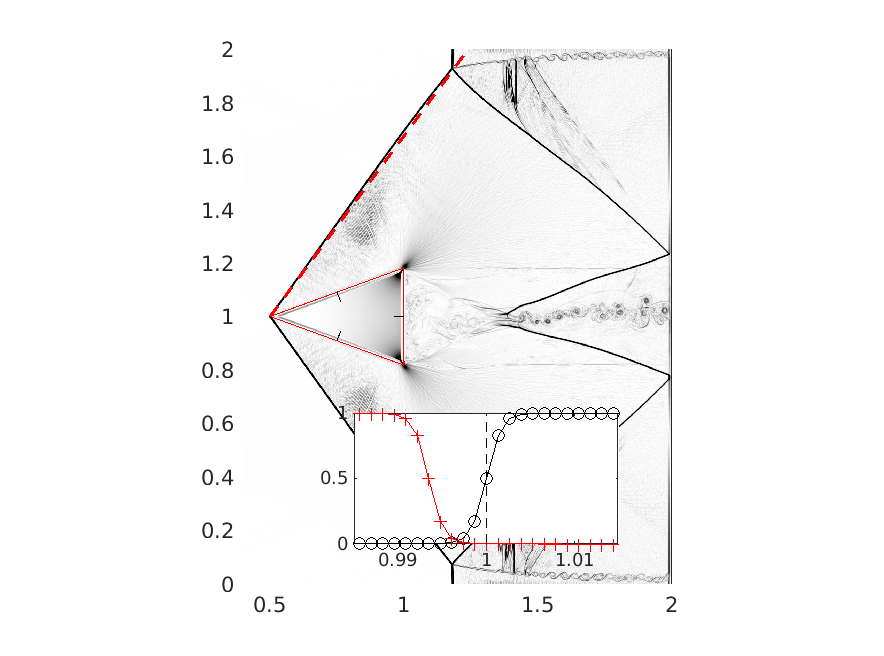}
	\includegraphics[width=.49\linewidth,viewport = 70 0 340 300,clip = true]{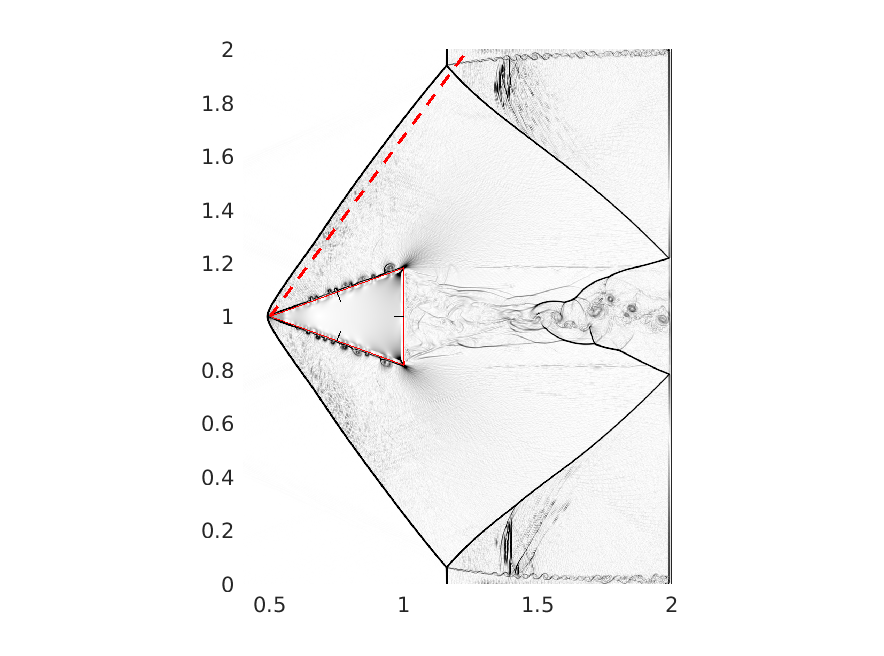}
	\caption{The modulus of the density gradient (pseudo-schlieren). Left: The slip boundary condition reproduces the shock location predicted by the (inviscid) gas dynamical theory. The triple points close to the upper and lower boundary arise from the periodic boundary conditions.
		\change{Inset: The reduced volume $\phi$ (o) and the normalized Darcy friction
			({\color{red}+}) around $x=1$ and along $y=1$ shows the smoothing of the boundary and the shift of the Darcy friction away from the boundary.
		} 
		Right: The non-slip boundary condition produces a lift-off and a slightly curved shock due to the flow displacement by the boundary layer, visible by the vortices at the wedge. 
	}
	\label{fig:wegde}
\end{figure}

A wedge in supersonic flow is now investigated. 
The case follows \cite{BoironChiavassaDonat2009} closely. 
Normalized quantities are used in this section. 
A two-dimensional domain of $2\times2$ with a wedge of half angel $\theta = 20^\circ$ and tip at $S= (0.5,1)$ is investigated. 
The main difference to the cited reference is a periodic boundary condition in $y$ for simplicity. 
Due to the hyperbolicity of supersonic flows, large parts of the flow can be directly compared. 
The in- and outflow is forced by a sponge term to Ma $= 2$, the reference values are $(\rho_0, u_0, v_0 , p_0 )= (1 , \mathrm{Ma} \sqrt{\gamma}, 0 ,1 ) $, $\gamma =1.4$. 

To handle the shocks, an adaptive filter as described above is used, where $\sigma$ is determined by a shock detector as described by \change{Bogey et al.} \cite{BogeyCacquerayBailly2009} and modified as in \cite{ReissSesterhenn2014} by a soft switch function. A threshold of $r_\mathrm{th}= 10^{-3}$ and a steepness of 
\change{$\lambda_F= 1/10$ was used, see appendix~\ref{app:filter} for more details. } 
%reference \cite{ReissSesterhenn2014} for a definition of the parameter. 

Two different set-ups are compared. A non-slip boundary condition which compares to \cite{BoironChiavassaDonat2009} and a slip condition
which can be directly compared with analytical predictions from gas dynamics, 
where inviscid flow and thereby non-slip conditions are assumed. 
In both cases, a resolution of $1536^2$ grid points as in \cite{BoironChiavassaDonat2009} is used.

Unsurprisingly, the slip boundary condition, made possible by the new method, agrees much better with the gas dynamical prediction, see Fig.~\ref{fig:wegde}.
\change{The  absence of numerical damping allows for fine details, but also yields high frequency noise, which is removed by filtering in post-processing.} 
The error for the non-slip condition depends on the resolution since the minimal boundary layer thickness (created by numerical dissipation) depends on it. 
For a slip boundary, the agreement is largely independent of the resolution. 
This has also practical consequences, especially in a first design phase, where a low resolution is often interesting when the boundary layer is not expected to be crucial for the design. 
Here, a lower resolution permits a shorter design process, so that the volume fraction approach might be of help.

\change{
\subsubsection{Wedge Shock wave interaction} 
	
Here, a shock-wedge interaction is discussed. 
The case follows a case presented by Kemm at al. \cite{KemmGaburroTheinDumbser2020} and  Dumbser et al. \cite{DumbserKaeserTitarevToro2007}.
A  a wedge of height $h=1$  and length $l=1$  is placed with its tip at (2,3) in a domain of size $(8\times 6)$.    	
A shock wave is located at $x=1$ at time $t=0$,  
with a  Mach number Ma$=1.3 $  enters a gas at rest ($u_0=0$) with $\rho_0 =1.4$ and $p_0=1$ with a adiabatic index of $\gamma=1.4$.  

\begin{figure}[t]
	\begin{center}
			\includegraphics[width=.45\linewidth,viewport = 10 0 400 300,clip = true]{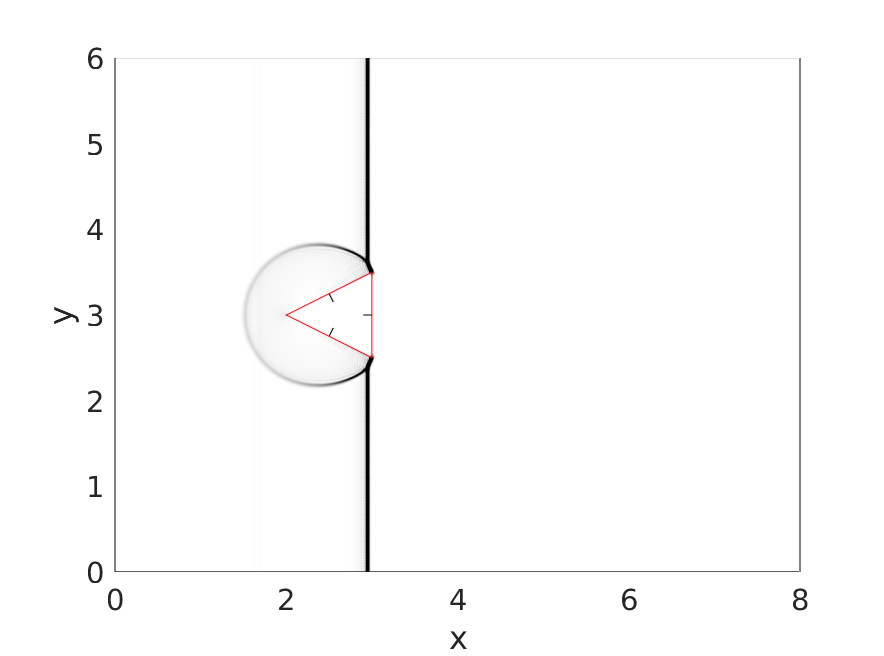}
	\includegraphics[width=.45\linewidth,viewport = 10 0 400 300,clip = true]{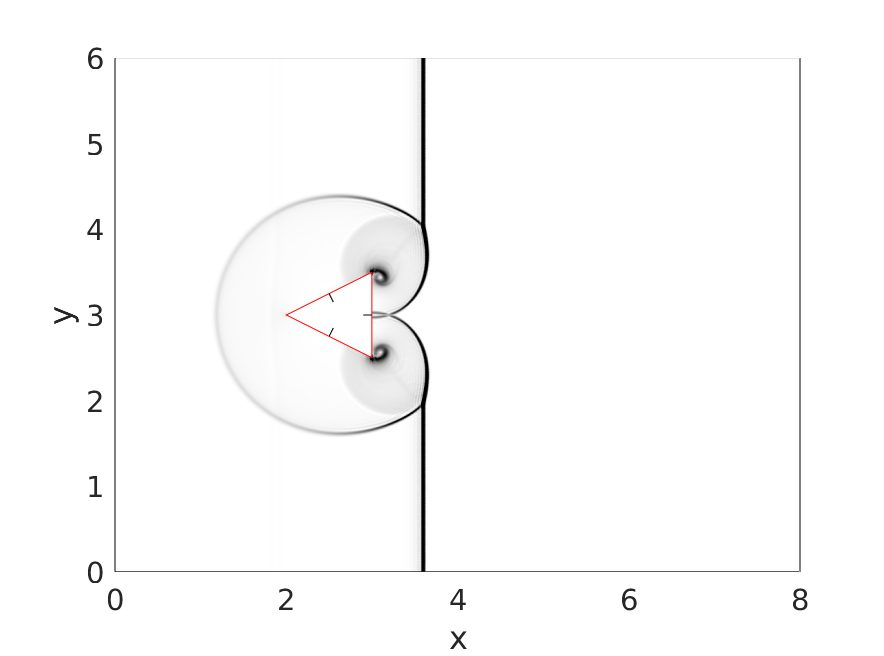}			
\end{center}

	\caption{The modulus of the density gradient (pseudo-schlieren) for the time $t=1.5$ and $t=2.0$.   
	}
	\label{fig:wegdeShock} % case wedgeShock/03 
\end{figure}

The domain is resolved with 1000$\times$750 points to be comparable with the additional degrees of freedom of discontinuous Galerkin elements of \cite{KapilaMenikoffBdzilSonStewart2001}. 
The reduced volume is again created from a signed-distance function and the smoothing by \eqref{boundFun} with $\delta= \Delta x $ with a value of $\eps= 10^{-8}$ in the object. 
 
The shock filter with $r_\mathrm{th}=10^{-4}$ and $\lambda_F=4$ is applied at every time step.  
Additionally, a global filter strength $\sigma_\mathrm{static} = 0.05$ is used to avoid noise in the simulation, see
appendix~\ref{app:filter} for details.  
The first/second-order stencil is used. 

The overall agreement is good.    
However, the shocks are not as sharp as in the reference simulation. 
This could be improved by a different shock detector (similar to a  flux limiter) 
or by optimized  filter coefficients. 
Artificial viscosity \cite{CookCabot2005} might be an option.  
This investigation is not in the scope of this publication.

}

\subsubsection{Pressure pulse in tube }

This example is motivated by a detonation wave in a tube entering a plenum of a turbine and aims at assessing the conservation quality. 
Such unsteady flows arise when coupling detonative and unsteady combustion with turbines to construct new types of gas turbines, see e.g. \cite{BengoecheaGrayReissMoeckPaschereitSesterhenn2018}. 
Here, high-pressure gradients appear, which challenges purely a Darcy-friction-based volume penalization, since it has to balance the high-pressure forces. Unlike a turbine, the domain is closed in the example, to allow a direct evaluation of conservation properties. 

The two-dimensional domain is 2m long and 1m wide, with a tube of diameter $0.2$m in the left part and a chamber in the right part. All walls have a thickness of 0.05m. 
Three NACA6412 profiles of cord length 0.3m tilted by $-(3/16)\pi$ are placed in the domain, mimicking a stator cascade, but mainly serving as non-trivial test objects with strong curvature and kinks.

\begin{figure}
	\begin{minipage}{.47\linewidth}
		\includegraphics[width=\linewidth,clip = true, viewport= 15 91 400 250]{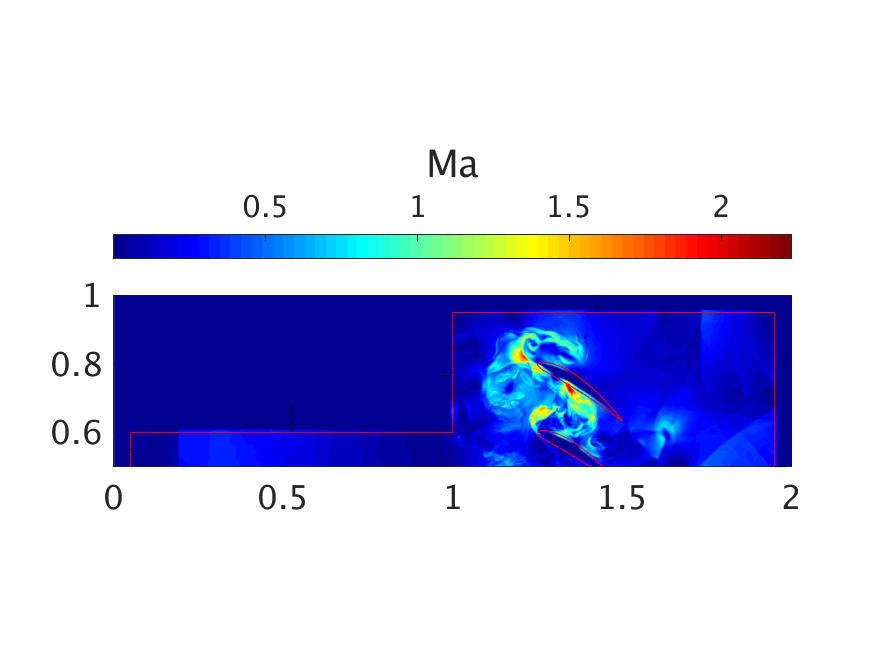}\\
		\includegraphics[width=.99\linewidth,clip = true, viewport= 15 60 400 239]{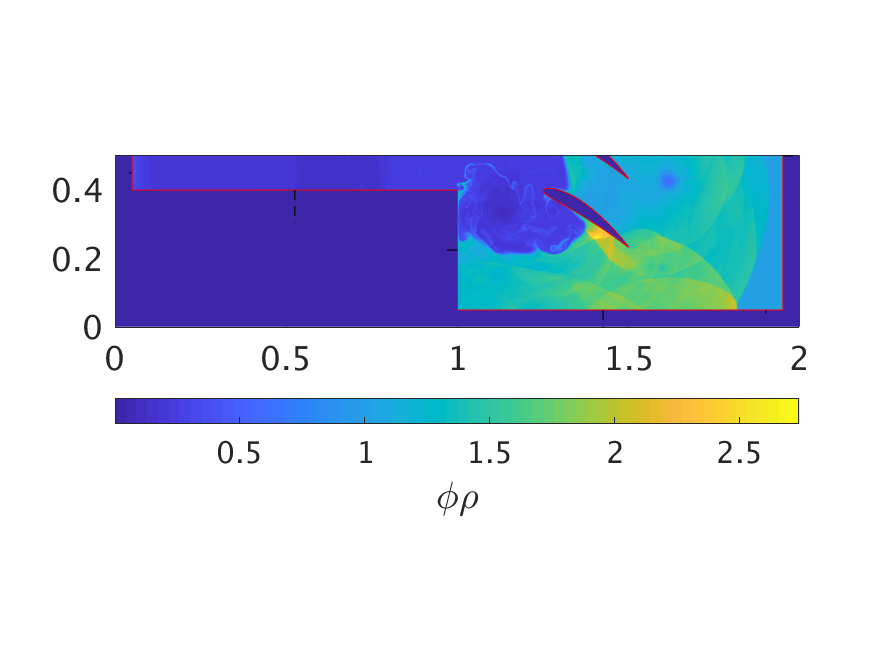}		 	
	\end{minipage}
	\begin{minipage}{.5\linewidth}
		\includegraphics[width=\linewidth,clip = true, viewport= 0 20 400 310 ]{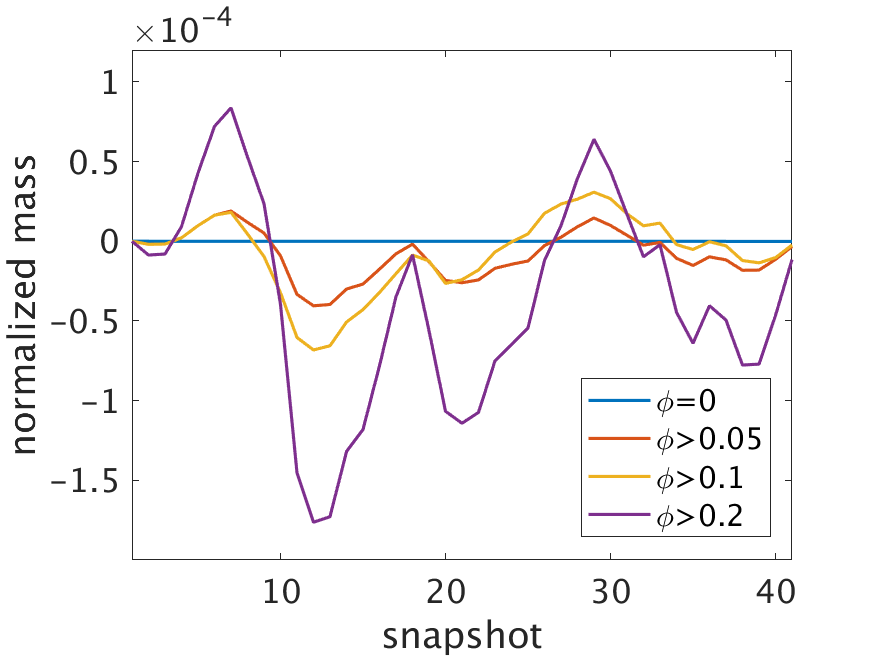}
	\end{minipage}
	\caption{Left: An unsteady jet flow entering into a closed chamber with obstacles is rich in physical phenomena and produces strong unsteady pressure forces on the boundaries. The Mach number is shown in the upper half and the volume fraction weighted mass density in the lower part. 
		Right: The conservation of the total mass depends on the definition of the integration domain, see text. }
	\label{fig:pulseJet}
\end{figure}

As start condition, an elevated temperature of $700$K and pressure of $4\cdot10^{5}$ is created in a patch between 
$x = $ [0.05m 0.8m] and $y=$ [0.2m 0.8m] smoothed by a hyperbolic tangent with a width of $\delta = 0.02m$. 
By this, the tube is filled nearly homogeneously to the point of $x=0.8$m. 
The increase of pressure and temperature inside the wall has negligible impact due to its small effective volume. 
Otherwise, we set $p_0 = 10^{5}$Pa, and $\rho_0 = 1$kg/m$^3$ and zero velocity everywhere. 
A CFL Number of $0.3$ is used, since otherwise non-smoothness in the flow (away from the walls) is created, yielding a timestep of $\Delta t \approx 2.67 \cdot 10^{-7}$. 

As before, a Darcy term with strength $1/(2 \Delta t)$ is used inside the wall with an offset of $7 \Delta x$ to create slip walls. 
A first/second-order filter is used which is applied everywhere in the object and at shock position in the flow. 
For this, the shock detector described for the wedge flow is used again, with $r_\mathrm{th}=10^{-5}$.

A supersonic jet ejects from the tube and decays before fully established, due to the small volume of the tube. 
By this, high-pressure loads and a large unsteadiness are created on the profiles and the walls. 
A snapshot of the Mach-number and the density is shown for the times $t=2.67$ms inf Fig. \ref{fig:pulseJet}. 
The total mass conservation is shown. 
If the density is integrated (summed-up over all grid points), near-perfect conservation is found, as expected from the conservative mass equation (\ref{mass}). 
Since we implement it in the form (\ref{massSkew}), a small error is introduced with an explicit time integration scheme (not strictly conserving quadratic invariants), which yields a relative error below $10^{-7}$ over the full simulation with $20000$ steps. 
To integrate the fluid domain, one has to choose the exact location of the wall, due to the ambiguity introduced by the smoothing of $\phi$. 
Here, this is done by choosing a threshold in $\phi$, thus defining a sharp mask. 
The larger the threshold is chosen, the more of the boundary region is omitted and the lower the conservation quality becomes, see Fig.~\ref{fig:pulseJet}. 
However, the conservation error fluctuates around the correct value, since the mass is not lost but just pressed into the wall smoothing region. 
Thus, this behavior might be better viewed as a slightly elastic wall than as a mass conservation defect. 
In practice, the integration of the full domain might be appropriate since the mass deep inside the object becomes as vanishingly small as $\phi$ does.  
Altogether, we can confirm good conservation.

\subsubsection{Vortex shedding} 

This case aims at a non-slip boundary and transformed grids. 
Transformed grids allow focusing on location with high gradients in the flow
and thereby reduce the total number of points and likewise the numerical effort. 
Since the volume fraction is part of the continuous equations with a simple physical interpretation, it is 
straightforward to generalize the scheme to transformed grids.

%Sr = 0.1727 , case 10 
%Sr = 0.1727 , case 11 

\begin{figure}[t]
\includegraphics[width=.6\linewidth,viewport = 0 30 400 290,clip = true]{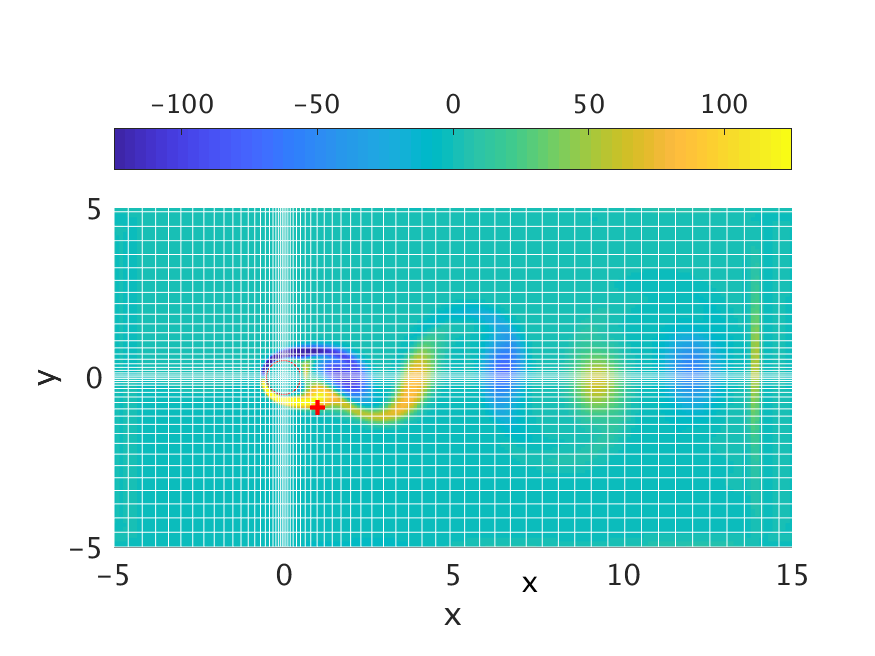}\quad
	\includegraphics[width=.33\linewidth,viewport = 140 10 400 280,clip = true]{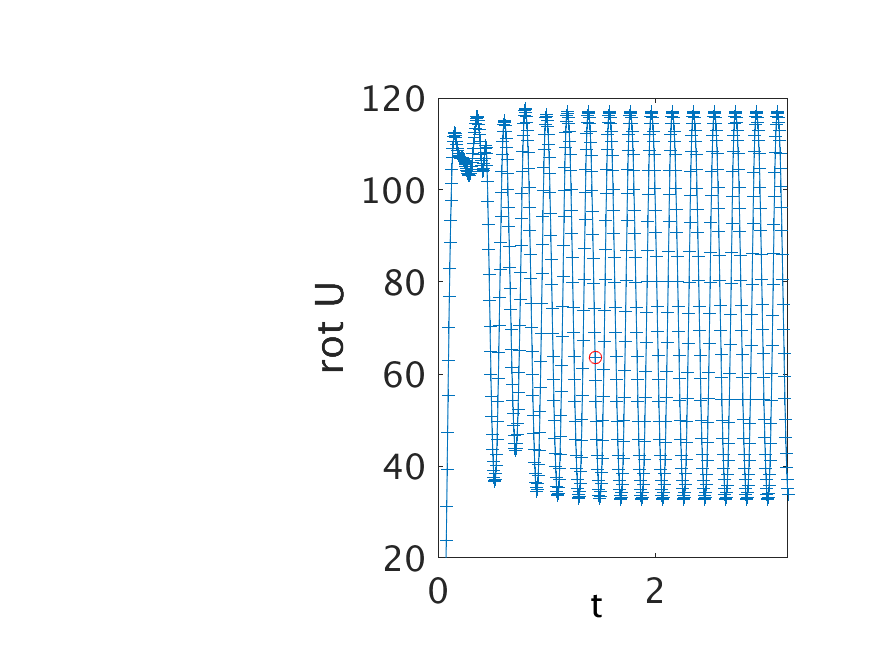}
	\caption{Left: A snapshot of the vorticity of the flow at time $t=1.44$. 
		Vorticity reference point as a red cross. The grid is shown as white lines, where only every fourth line is plotted. 
	  Right: The vorticity at the reference point as a function of time shows a Strouhal number of Sr=0.173.	}
	\label{fig:vortexStreet}
	% case: immersedObjects/cylinderVortexStreet/11
\end{figure}

A flow with Reynolds number Re$=(u_0 D)/\nu= 150$ around a cylinder with diameter $D=1$ is simulated in a domain of size ($20\times15$), using normalized quantities in this section. The in-flow velocity is $u_0=30$ with a speed of sound $c_0=374$ yielding a Mach-number of Ma=0.08 so that it can be safely regarded as incompressible. 

The transformed grid with $N_\xi\times N_\eta = (256 \times 166) $ is given by 
\change{
\begin{align}
x &= (\tilde x - \tilde x_1)/(\tilde x_{N_\xi} - \tilde x_1)\cdot L_\xi  \\    
y &= (\tilde y - \tilde y_1)/(\tilde y_{N_\eta} - \tilde y_1)\cdot L_\eta 
\end{align}
where 
\begin{align}
\tilde x  & = \xi - \alpha(\tanh( (\xi - (x_0+\alpha + x_\delta  ) )/\psi ) +1  )  \\ 
 \tilde y &  = \eta- \alpha(\tanh( (\eta- (y_0+\alpha  ) )/\psi ) ) -\alpha .
\end{align}
Here, $\xi  = (L_\xi +2\alpha )(0,\dots,N_\xi-1 )/( N_\xi-1)$ 
and 
$\eta = (L_\eta+2\alpha )(0,\dots,N_\eta-1)/(N_\eta-1)  $. 
The parameters are 
$\alpha = 7.2$ and $\psi  =8$ and $x_\delta = -0.55$
From this, the coordinates are created by normalizing to the desired domain size. 
}
The resulting grid is shown in Fig.~\ref{fig:vortexStreet}, where every fourth grid line is plotted. 
A CFL number of $1/2$ was used. 

The grid transformation is introduced by rewriting divergence and gradient operators with the help of local bases, \change{ as discussed in the appendix~\ref{app:detail}.} 
In the end, all divergence operators are replaced by the so-called conservative form and the gradient operators by the so-called non-conservative form, yielding a discretely consistent (and conservative) scheme, if all derivatives are replaced by finite difference approximations with a (skew-)symmetric stencil. 
Using summation by part operators results in well-defined fluxes at the boundaries result, see \cite{ReissSesterhenn2014} for details\footnote{In the cited reference, the conservative form is used for the pressure terms. However, using the non-conservative form is more consistent with the incompressible theory \cite{Reiss2015}.}. The grid is orthogonal; a non-orthogonal grid can also be handled by this approach, it was however not explicitly tested. 

The initial flow is the potential flow discussed \change{ in Sec.~\ref{sec:potFlow}} with an additional Gaussian velocity defect at $(x_0=0,y_0=1/2)$, with a width of $\sigma_\mathrm{disturb}  = 1/2$ and an amplitude $-20$ to trigger the vortex shedding: \change{$u_0= u_\mathrm{pot} + (-20)\exp\left(-((X-x_0)^2 + (Y-y_0)^2)/\sigma_\mathrm{disturb}^2  \right) $}. 
The inflow and outflow are created by sponge terms enforcing the potential flow around a cylinder, where a quadratic sponge of thickness $10 L_\xi/N_\xi$ is used. 
The sponge is visible by creating extra vorticity close to the outflow in Fig.~\ref{fig:vortexStreet}. 
A non-reflecting boundary condition with reference conditions as the initial condition is used at the top and bottom. 

The described filter of second/fourth-order is used inside the cylinder, with a small amplitude of $1/100$ in the whole domain to avoid grid to grid oscillations, presumably created by the boundary conditions. 
The \change{friction-like effect} of the filter is by this small by the order 1/20 compared with the physical friction and should, thus, have little influence on the Reynolds number. 

The expected vortex shedding is found after a short transition time with a Strouhal number of $ Sr =  0.173$, which is in agreement with the literature. It is found to be robust towards simulation details like resolution.
The smoothing of $\phi$ of $\delta = (L_\xi/N_\xi)/2$ was used, which is close to one grid spacing at the cylinder wall due to the grid refinement. A sharper smoothing for the Darcy term was used with $\delta_\mathrm{Darcy}= \delta/2$, to ensure a quick decay of the tangential velocity, however, small sensitivity was found in these parameters. 
This is important since the good slip conditions possible by the volume fraction $\phi$ might be taken as an indicator of possible problems with the non-slip conditions. No signs for this are, however, found in the presented investigations. 

Altogether, we find satisfactory results for this low Mach number, viscosity-driven case.

\section{Conclusion and outlook}
\label{sec:conclusion}
A method to describe immersed objects was presented. 
It is easy to implement and has a clear physical interpretation. 
Furthermore, it has good conservation properties. 
It naturally allows implementing slip boundaries and adiabatic boundary conditions. 
The non-penetration condition can be implemented with only a minor increase in the stiffness of the equation and 
near-perfect pressure-tight boundaries are possible, even for high-pressure gradients. 

The derivation follows considerations of \change{Kevlahan et al.} \cite{KevlahanDubosAechtner2015} for shallow water equation and builds on a physical consistent treatment of the porosity, understood as a reduced volume fraction available for the fluid. 
By this, it becomes equivalent to the available cross-section. 

\change{
	While the main method works well, the filter, needed especially for the treatment of shocks needs manual adjustment and does not produce  as sharp shocks as shock limiter  methods \cite{KemmGaburroTheinDumbser2020}. 
	This will be addressed in future works by optimizing the filter for the specific discretization of the Euler equations.    
}

The method does not pose any extra difficulties in a parallel environment. 
For more complex geometries large simulations in 3D are in preparation. 
To avoid the increased stiffness for very sharp boundary functions, a better resolution close to the boundary is favorable, if such a high localization of the wall is demanded by the problem. 
Alternatively, to the shown local grid transformations, a local grid refinement as in \cite{EngelsSchneiderReissFarge2019} can be used, which is also under investigation. 
The possible non-slip conditions and the very good reflectivity make this equation a good candidate especially for aero-acoustic problems, which are currently further investigated. 
%\smallskip 

%\bibliography{local}{}
%\bibliographystyle{plain}

\subsection*{Acknowledgements}
I thank Thomas Engels and 
Nicholas Kevlahan for valuable discussions. 
%\end{acknowledgements}
% Authors must disclose all relationships or interests that 
% could have direct or potential influence or impart bias on 
% the work: 
%
 \section*{Conflict of interest}
 The author declares that he has no conflict of interest.
 
 \section*{Availability of data}
All data is created by the described method and settings. 
Simulation data sets are available from the corresponding author on reasonable request. 

\section*{Funding}
Funded by the Deutsche Forschungsgemeinschaft (DFG, German Research Foun-dation) - Projektnummer 200291049 - SFB 1029 (A04)

\bibliography{local}  % name your BibTeX data base

\bibliographystyle{plain}       % APS-like style for physics

\appendix

\section{Numerical Details}
 \label{app:detail}
 
Here, details of numerical implementation are described. 

The time stepping consists of an (explicit) Runge-Kutta-4, which evaluates as the spatial part of the equations discretized by (explicit) finite differences. 
A filter is used after the time-stepping to remove oscillatory behavior.  
The main program structure is presented in algorithm~1. 
\begin{algorithm}
	%\label{tab:main}
	\caption{ The main solver  }
	\begin{algorithmic}
	\label{alg:J2Grad}
		 
		\STATE{ $q_0$ = initialFlow() } \hfill \# {inital condition for the case} 
		\STATE
		
		\FOR{$n=1$ \TO $N_{\mathrm{timeSteps}}$ } 
		\STATE
		\STATE { \#  Runge Kutta 4 time stepping:} 		
		\STATE  \quad {$k_1$ := {rhs}($q_0$)    } %\hfill \COMMENT{SVD with $r_k$ modes}
		\STATE  \quad{$k_2$ := {rhs}($q_0 + \frac{\Delta t}2 k_1 $)    } %\hfill \COMMENT{SVD with $r_k$ modes}
		\STATE  \quad{$k_3$ := {rhs}($q_0 + \frac{\Delta t}2 k_2 $)    } %\hfill \COMMENT{SVD with $r_k$ modes}
		\STATE  \quad{$k_4$ := {rhs}($q_0 +  {\Delta t}\,  k_3 $)    } %\hfill \COMMENT{SVD with $r_k$ modes}
     
		\STATE  {$q_1 := q_0 + \Delta t (k_1 + 2 k_2 +2 k_3+k_4)/6 $ } 
			
		\STATE  % $n_k = | q^k|_F $ 					\hfill \COMMENT{Frobenius Norm}
		
		\STATE { \# filter if demanded } 		

		\IF {filterAtStep(n)}
		    \STATE{$q_c$ = transformToConservative($q_1$)}  \hfill \# { conservative variables $\rho, \rho u_\alpha, \rho e_t$  } 
		    \STATE{ $\sigma_\alpha$ = calcFilterStrength($q_1$) \hfill \# {adaptive filter strength for shocks}   }
		    \FOR{$\alpha=1$ \TO spatialDimensions  } 
			\STATE {$   q_c    =  q_c  +   (\bar D_\alpha  (M (\sigma_\alpha\phi) ) D_\alpha q_c)/\phi    $ }   	   	
			    \ENDFOR
			 \STATE{$q_1$ = transformFromConservative($q_c$)}  \hfill \# { back to calculation variables } 
		\ENDIF
		
		\STATE { \# plot or save state} 		
		
		\STATE {$q_0 := q_1$ \hfill \# prepare next time step } 
		
		\ENDFOR

		\RETURN  
	\end{algorithmic}
\end{algorithm}

The right-hand side (rhs) is the spatial part of the equations.
The flow is described in the variables are $\sqrho, (\sqrho u_\alpha)$ and $p$. All other quantities 
are expressed by theses, e.g. $ u_\alpha = (\sqrho u_\alpha)/\sqrho$. 
The temperature  is calculated from the ideal gas law $T= p/((\sqrho)^2 R_s )$ with the specific gas constant 
$ R_s= R/W$ calculated from the universal gas constant $R$ and the molecular weight $W$.     
For transformed grids and temporal constant reduced volume $\phi$ the discretized equations are 
\begin{eqnarray}
	J \phi  2\sqrho \partial_t  \sqrho +      
	D_{\xi^\beta}  (\tilde  u_\beta      \phi \rho)   &=&0 \label{mass2Ddisc}\\[0pt]
	J \phi  \sqrt{\rho} \partial_t   (\sqrt{\rho}  u_\alpha ) 
	+ 
	\halb (D_{\xi^\beta}  \tilde  u_\beta \rho \phi u_\alpha + \phi  \rho \tilde  u_\beta D_{\xi^\beta}   u_\alpha) % D^{ u\rho } 
	+   
	\phi \bar D_{x_\alpha}  p
	&=& \no\\
	\phi  \chi  (u_\alpha^\mathrm{t} - u_\alpha  )  
	+ D_{\xi^\beta} \phi\tilde  \tau_{\alpha\beta}  \label{momUalphadisc}&&\\  
	J\phi \frac{1}{\gamma-1} \partial_t  p 
	+ 
	\frac {\gamma}{\gamma-1}  D_{\xi^\beta} \tilde  u_\beta \phi  p 
	- 
	u_\alpha \phi \bar D_{x_\alpha}   p
	&=& \no\\  
	- u_\alpha    
	[ \phi  \chi  (u_\alpha^\mathrm{t} - u_\alpha  ) 
	+D_{\xi^\alpha} \phi \tilde \tau_{\alpha\beta} ] 
	+D_{\xi^\alpha} u_\beta \phi\tilde \tau_{\alpha \beta}  - D_{\xi^\alpha} \tilde\varphi_\alpha %\label{p3Dskew}
	&&\label{en2Ddisc} .
\end{eqnarray}
From theses equations the time derivatives of the calculation variables directly follows, yielding the structure 
$\partial_t q = \mathrm{rhs}(q) $:
\begin{eqnarray}
	&& \partial_t  \sqrho  = -\left[       
	D_{\xi^\beta}  (\tilde  u_\beta      \phi \rho) \right]/(J \phi  2\sqrho)   \\  % &=&0 \label{mass2Ddisc}\\[0pt]
	&&\partial_t   (\sqrt{\rho}  u_\alpha ) = \\ 
	&& \qquad\no- \left[ 
	+ 
	\halb (D_{\xi^\beta}  \tilde  u_\beta \rho \phi u_\alpha + \phi  \rho \tilde  u_\beta D_{\xi^\beta}   u_\alpha) %  D^{  u\rho }  
	+   
	\phi \bar D_{x_\alpha}  p
	-
	\phi  \chi  (u_\alpha^\mathrm{t} - u_\alpha  )  
	- D_{\xi^\beta} \phi\tilde  \tau_{\alpha\beta}\right]/(J \phi  \sqrt{\rho} )     \\ %\label{momUalphadisc}\\  
	&&\partial_t  p 
	= \\&&
	\qquad-\Bigl[ 
	\frac {\gamma}{\gamma-1}  D_{\xi^\beta} \tilde  u_\beta \phi  p 
	- 
	u_\alpha \phi \bar D_{x_\alpha}   p  +  %\\ \no 
	 u_\alpha    
	[ \phi  \chi  (u_\alpha^\mathrm{t} - u_\alpha  ) 
	+D_{\xi^\alpha} \phi \tilde \tau_{\alpha\beta} ] \\ \no &&
	 \qquad \qquad -D_{\xi^\alpha} u_\beta \phi\tilde \tau_{\alpha \beta}  + D_{\xi^\alpha} \tilde\varphi_\alpha \Bigr]\frac{\gamma-1}{J\phi}   %\label{p3Dskew}
	%\label{en2Ddisc} .
\end{eqnarray}
All field quantities are assumed to be multiplied or divided pointwise.
The heat flux $\tilde \varphi_\alpha  $ was introduced, which is not to be confused with the reduced volume.  
For Cartesian grids $J\equiv1$, $ \tilde u_\alpha = u_\alpha$ and $ \bar D_{x_\alpha} =  D_{x_\alpha} $.
The derivative matrices are central finite difference derivative matrices % $ D_{\xi^\alpha}\equiv D_{\xi^\alpha} $
 along one spatial direction, i.e., one index of the 
discretized field. In detail, assuming summing convention: 
$$ 
(D_{\xi^1} f)_{i,j,k} =  (D_{\xi^1})_{i,i'} f_{i',j,k} \quad    
(D_{\xi^2} f)_{i,j,k} =  (D_{\xi^2})_{j,j'} f_{i,j',k} \quad   
(D_{\xi^3} f)_{i,j,k} =  (D_{\xi^3})_{k,k'} f_{i,j,k'}   
$$  
A curvilinear grid is introduced by a transformation following Thompson et al. \cite{ThompsonWarsiMastin1985}. 
The transformation is implied by the discrete values of the coordinates at the grid points 
$$
r= 
\left(
\begin{matrix}
	x_1 \\
	x_2\\
	x_3
\end{matrix}
\right)
\equiv 
\left(
\begin{matrix}
x_1(\xi_1,\xi_2,\xi_3) \\
x_2(\xi_1,\xi_2,\xi_3)\\
x_3(\xi_1,\xi_2,\xi_3)
\end{matrix}
\right).
$$
where the calculation space variables are discretized equidistantly
$(\xi_1)_i= i \Delta \xi_1,(\xi_2)_j= j\Delta \xi_2,(\xi_3)_k= k \Delta \xi_3 $. 
From these grid points the local base vectors 
\begin{eqnarray}
	{\bf e}_\alpha= D_{\xi^\alpha} {\bf r} 
\end{eqnarray}
and the matrix containing the metric factors\footnote{
	Note that for general transformations in three dimensions a rewriting of these factors is necessary to preserve consistencies \cite{ThomasLombard1979}. 
	 This modification can directly be included in the here presented method.  } 
 are calculated
 %, where in the second step the transformation is restricted to two dimensions, used for all examples in this report:
\begin{eqnarray}
	\mathbf{m} = 
	\left(  \mathbf{e}_2\times \mathbf{e}_3   ,\; 
%\mathbf{m_2} = 
\mathbf{e}_3 \times \mathbf{e}_1   ,\; 
%\mathbf{m_3} = 
\mathbf{e}_1 \times \mathbf{e}_2  
\right) 
= 
\left(\begin{matrix}
	D_{\xi^2} x_2 & - D_{\xi^1} x_2 & 0 \\
	- D_{\xi^2} x_1 & D_{\xi^1} x_1 & 0 \\
	0             & 0             & J \\
\end{matrix} 
\right) .
\end{eqnarray}
In the second step a two-dimensional transformation ($x_1 = x_1(\xi_1,\xi_2)$, $x_2 = x_2(\xi_1,\xi_2)$, $x_3= \xi_3$) was taken, which is used  in this report; in this case 
all fields are assumed to be independent of $\xi_3$. 
The Jacobian is $J=({\bf e}_1\times {\bf e}_2)\cdot {\bf e}_3$. 
With this, the derivatives in physical space ($x$) can be rewritten in computational space ($\xi$) as 
$$
D_{x^\alpha} f =  \frac 1 J \mathbf{m}_{\alpha \beta}  D_{\xi^\beta}  f 
$$
or 
$$
D_{x^\alpha} f = \frac 1 J   D_{\xi^\beta} \mathbf{m}_{\alpha \beta} f
$$
which is called the non-conservative and the conservative form.
Despite this naming conservative schemes can be constructed from both forms, see also \cite{Reiss2015} on this point. 
They are analytical identical but differ in general for a discretized derivative. 
These two forms allows to formulate the equations (\ref{massSkew}-\ref{energySkew}) on curvilinear grids. 
%In places where the second form is used 
The metric can mostly be absorbed into the effective velocity
$ 	\tilde u_\alpha = \mathbf{m_{\beta\alpha}}  u_\beta $. 
The pressure gradient is written as
$
\bar D_{x_\alpha}  = \mathbf{m}_{\alpha\beta}  D_{\xi^\beta}. 
$
 The heat flux becomes by the combination of the non-conservative for the inner derivative, and the conservative form for the outer derivative a symmetric operator 
$\tilde \varphi_\alpha = -\lambda \frac \Phi J  \mathbf{m}_{\alpha\gamma} \mathbf{m}_{\alpha\beta}    D_{\xi^\beta} T $. 
In the same manner, the shear viscosity is found to be 
\begin{eqnarray}
	 \tilde \tau_{\alpha\beta} = \mu \left( 
	  \frac {\mathbf{m}_{\delta\beta}\mathbf{m}_{\alpha\gamma}}{J}  D_{\xi^\gamma} u_\delta +
	  \frac {\mathbf{m}_{\delta\beta}\mathbf{m}_{\delta\gamma}}{J}  D_{\xi^\gamma} u_\alpha  
	 \right) 
	 +
	 (\mu_d   -2/3 \mu)  
	 \delta_{\alpha,\omega}   
	  \frac {\mathbf{m}_{\beta\omega} \mathbf{m}_{\gamma\delta}}{J}   D_{\xi^\delta}  u_\gamma. 
\end{eqnarray}
In this form, the spatial part is fully explicit. 

\section{Conservation} 	
\label{app:cons}

The conservation properties of the scheme build on the conservation properties of the spatial and the temporal discretization. 
While the spatial part is strictly conservative, the temporal discretization used in this publication produces a small error, so that strict conservation is assured only for $\Delta t \to 0$.
Special implicit methods to overcome this are discussed below. 
\smallskip 

The show the spatial conservation consider the equations (\ref{mass2Ddisc}-\ref{en2Ddisc}). 
We assume periodic grids in the following so that the discrete derivatives fulfill the telescoping sum property 
\begin{eqnarray}
 \mathbf{1}^T D_{\xi^\alpha} f   \equiv  \sum_{i,j,k} 1 (D_{\xi^\alpha} f)_{i,j,k}  = 0  
 \label{telescope} 
\end{eqnarray}
and  skew symmetry
\begin{eqnarray} 
  D^T_{\xi^\alpha}   = -  D_{\xi^\alpha} f ,
\end{eqnarray} 
 as a result of the skew-symmetry of the one-dimensional (central) derivative. 
The non-periodic case can be treated with SBP matrices in a fully consistent manner (as in \cite{ReissSesterhenn2014}), but is not detailed here for the sake of brevity.  
Viscous terms are omitted in the discussion  but can be included in a straightforward manner. 
 
The conservation of the total mass $M$  follows directly by summing the equation $ \mathbf{1}^T ( \ref{mass2Ddisc})$ from which the spatial part evaluates to zero due to \eqref{telescope} and the temporal part becomes with the product rule\footnote{ 
Assuming here a time constant $J$ and $\phi$. Time-dependent grids are not considered in this report, while a time-dependent reduced volume yields the extra term demanded by the mass conservation, as seen from \eqref{massMoving}.} 
\begin{eqnarray}
 \mathbf{1}^T J \phi  2\sqrho \partial_t  \sqrho  
 +
  \mathbf{1}^T D_{\xi^\beta}  (\tilde  u_\beta      \phi \rho) 
 =  \partial_t  (\mathbf{1}^T J \phi     (\sqrho)^2) = \partial_t M = 0 .
\end{eqnarray}  
 The change of total momentum is calculated from $ \frac 1 2   u_\alpha^T \eqref{mass2Ddisc} + \eqref{momUalphadisc} $:
\begin{eqnarray}
  u_\alpha^T	J \phi  \sqrho \partial_t  \sqrho +      
u_\alpha^T  D_{\xi^\beta}  (\tilde  u_\beta      \rho \phi )/2   +&& \no \\[0pt]
\mathbf{1}^T J \phi  \sqrt{\rho} \partial_t   (\sqrt{\rho}  u_\alpha ) 
 + 
 \halb 
 ( \mathbf{1}^T D_{\xi^\beta} ( \tilde  u_\beta \rho \phi)   u_\alpha 
 + 
 \mathbf{1}^T (\phi  \rho  \tilde  u_\beta) D_{\xi^\beta}   u_\alpha) % D^{ u\rho }  
 +   
 \mathbf{1}^T \phi \bar D_{x_\alpha}  p  &&  \no \\
 = 
  u_\alpha^T	J \phi  \sqrho \partial_t  \sqrho +    \mathbf{1}^T J \phi  \sqrt{\rho} \partial_t   (\sqrt{\rho}  u_\alpha )  
  + 
   \mathbf{1}^T \phi \bar D_{x_\alpha}  p \no\\
=    
  \partial_t  \mathbf{1}^T J \phi  \sqrt{\rho} (\sqrt{\rho}  u_\alpha )  
+ 
 \mathbf{1}^T\phi  \bar D_{x_\alpha}  p  =	\mathbf{1}^T \phi  \chi  (u_\alpha^\mathrm{t} - u_\alpha  )   \label{momConsDisc}
\end{eqnarray}     
 The telescoping sum property in the first term of the momentum transport and the combination 
\begin{eqnarray}
 u_\alpha^T  D_{\xi^\beta}  (\tilde  u_\beta      \rho \phi ) 
 +
  \mathbf{1}^T (\phi  \rho  \tilde  u_\beta) D_{\xi^\beta}   u_\alpha 
  = 
 u_\alpha^T  D_{\xi^\beta}  (\tilde  u_\beta      \rho \phi ) 
+
 (\phi  \rho  \tilde  u_\beta)^T  D_{\xi^\beta}   u_\alpha    = 0 
\end{eqnarray}
becomes zero by the skew-symmetry of the discrete derivative since 
$  (\phi  \rho  \tilde  u_\beta)^T  D_{\xi^\beta}   u_\alpha =  
\left( (\phi  \rho  \tilde  u_\beta)^T  D_{\xi^\beta}   u_\alpha\right)^T 
= 
u_\alpha^T D_{\xi^\beta}^T (\phi  \rho  \tilde  u_\beta) = - u_\alpha^T D_{\xi^\beta} (\phi  \rho  \tilde  u_\beta)
  $. 
  The first term in \eqref{momConsDisc} is the temporal change of the total momentum the second is the source expected due to reduced effective volume by the embedded objects; 
  it is a discrete form of the integral 
  $\int_\Omega p\, \partial_{x_\alpha}  \phi  \sim 
  \mathbf{1}^T  \bar D_{x_\alpha}  \phi  p  
  -  \mathbf{1}^T\phi  \bar D_{x_\alpha}  p  $, since the first term is zero by the telescoping sum property. The term proportional to $\chi$ is the momentum source due to the Darcy friction term. All momentum changes are thereby a direct consequence of the source terms modeling the objects.  
  
  The energy conservation is obtained by the combination of the equation of internal energy and kinetic energy. 
  The latter is calculated by $u_\alpha^T \eqref{momUalphadisc}$:
 \begin{eqnarray}
 (u_\alpha \sqrt{\rho})^T  	J \phi   \partial_t   (\sqrt{\rho}  u_\alpha ) 
 + 
 \halb u_\alpha^T (D_{\xi^\beta}  \tilde  u_\beta \rho \phi u_\alpha + \phi  \rho \tilde  u_\beta D_{\xi^\beta}   u_\alpha) % D^{  u\rho }  
 +   
u_\alpha^T \phi \bar D_{x_\alpha}  p &&\no\\
 =   u_\alpha^T \phi  \chi  (u_\alpha^\mathrm{t} - u_\alpha  )&&
 \label{kinEnergyDisc}
  \end{eqnarray}
 The second term, the transport,  can be regarded as a quadratic form $u_\alpha^T D^{\rho \phi \mathbf{u}} u_\alpha  $ over the skew symmetric matrix $  D^{\rho \phi \mathbf{u}} =(D_{\xi^\beta}  \{\tilde  u_\beta \rho \phi\}  + \{\phi  \rho \tilde  u_\beta\} D_{\xi^\beta}  ) = -(D^{\rho \phi \mathbf{u}})^T $, which is zero in general. 
 The skew symmetry becomes clear if one considers the pointwise multiplications in the operator as diagonal matrix $\{\phi  \rho \tilde  u_\beta\}  $ with the corresponding values on the diagonal. 
 This skew-symmetry is the central point of the method as it avoids any numerical change of the kinetic energy. 
 In most schemes, the momentum equation can change the kinetic energy, which often demands the introduction of numerical dissipation even for smooth flow fields for stability reasons.  
 
 Calculating the internal energy part from $\mathbf{1}^T \eqref{en2Ddisc}  $, where the divergence terms again evaluate to zero, combines with \eqref{kinEnergyDisc} to yield 
 \begin{eqnarray}
	  \partial_t 
	  \left[ \mathbf{1}^T  \frac{ J\phi  p }{\gamma-1}  	
	  + 
	  \halb \partial_t  ( u_\alpha \sqrt{\rho})^T  	J \phi     (\sqrt{\rho}  u_\alpha ) 
	  \right]
&=& 0.  	
 \end{eqnarray}
 The pressure terms in kinetic and internal energy couple the kinetic and internal energy and cancel, as in the analytical case. 
 The Darcy term cancels if it is included in the internal energy equation \eqref{en2Ddisc}. 
 Since the change of kinetic energy by the Darcy term in the momentum equation might be compensated by the object, it is justified modeling that the term proportional to $\chi$ is omitted in   \eqref{en2Ddisc} and the total energy is not constant. 
 This can be interpreted as a penalization term for the energy as used in \cite{BoironChiavassaDonat2009} and is done in the simulations presented in this report. 
 The term  could easily  be included if strict energy conservation is desired. 
 
 The boundary conditions at the domain boundary are either periodic boundaries, implemented by a periodic derivative matrix in this direction, or are created by quadratic sponges as discussed in \cite{Mani2012} or are characteristic boundaries where a decoupling in Riemann waves is done with respect to a reference state and the in-going waves are set to zero, as discussed in the appendix of \cite{HossbachLemkeReiss2021} with a system matrix in a direction perpendicular to the boundary, see \cite[sec 16.5]{Hirsch1990} for details.  
 
 \medskip 
 
 The product rule, which was used for the time derivative,  is violated in most time discretizations.
 The conservation of quadratic invariants, needed in this approach for strict conservation, are respected by the implicit Gauss-Lobatto methods \cite{BrouwerReissSesterhenn2014b}. 
 The special choice of variables allows to use these methods and directly guarantees strict conservation, however, the error with the RK4 is so small that the numerical effort for the implicit time integration does not seem justified in most cases \cite{BrouwerReissSesterhenn2014b}. 
 For example, the relative error of the conservation mass in the full domain of the pulse tube test case was below $10^{-7}$ for the mass, which seems sufficient for the majority of practical studies.

\section{Filter details} 
\label{app:filter}
Numerical dissipation is introduced by a filter in the presented report. 
It ensures a sufficient amount of dissipation for shocks and allows to suppress highly oscillatory behavior, which can be created by the nearly non-smooth $\phi$ within the objects by waves entering from the physical domain. 
The basic properties of the filter are discussed in section \ref{sec:filter}.

Here, details of the shock detector are provided, which is a slightly modified version of the detector discussed in \cite{BogeyCacquerayBailly2009}.  
The core idea is to create a shock detector to locally activate filtering.
The filter can also be activated within the objects or (e.g. with a small strength) in the full region.

 The shock detection is based on the local divergence 
\begin{eqnarray}		
	\Theta = \mathrm{div}(\phi \mathbf(u)) ,
\end{eqnarray}
	from which the non-smoothness is calculated by an expression similar to a second derivative 
\begin{eqnarray}		
	 (D_{\xi_\alpha}  \Theta)_{i,j,k} = (-\Theta_{i+1,j.k} + 2 \Theta_{i+1,j,k} - \Theta_{i-1,j,k} )/4 ,
\end{eqnarray}
from which a magnitude is calculated by 
\begin{eqnarray}
(D_{\xi_\alpha}  \Theta)_{i,j,k}^\mathrm{magn} = 
\frac 1 2 \left[ 
(D \Theta_{i,j,k} -D \Theta_{i+1,j,k})^2 
+
(D \Theta_{i,j,k} -D \Theta_{i-1,j,k})^2 
\right], 
\end{eqnarray}
and likewise in the other spatial directions. 
Other non-smoothness criteria are possible for example, a adaptive filter based on the WENO detector is presented in 
\cite{VisbalGaitonde2005}, different detectors are compared in \cite{Pirozzoli2011a}. 
The non-smoothness parameter $ (D_{\xi_\alpha}   \Theta)^\mathrm{magn} $ is compared with the grid spacing and the local speed of sound 
$ 
r_{\alpha} = \frac{ (D_{\xi_\alpha}  \Theta)^\mathrm{magn} }{ c^2/\Delta \xi_\alpha^2}  
$ 
from which finally the filter strength is calculated as
\begin{eqnarray}
\sigma^\mathrm{shock}_\alpha   = 1-\tanh\left(\frac{r_\mathrm{th}}{r_{\alpha}\lambda_F}\right). 
\label{sigmaShock}
\end{eqnarray}
Here, $r_\mathrm{th}$ and $ \lambda_F$ are adjustable parameters. 
The procedure is similar to a flux limiter, which create in a non-linear manner a local amount of dissipation. 
Flux limiters usually go from high order schemes to low order up-wind methods, while here a purely dissipative term is added.  
The functional form \eqref{sigmaShock} is different from \cite{BogeyCacquerayBailly2009}, where a non-smooth function is used, creating a by switching sometimes high-frequency noise in the solution. 
This adaptive part can be combined with a user-defined static (and locally varying) filtering strength by 
$ 
\sigma_\alpha  = \max (\sigma^\mathrm{shock}_\alpha, \sigma_\mathrm{static}) 
$. 
The static part is for example a filter acting within the solid objects. 
The construction ensures that $0\le \sigma_\alpha\le 1$, ensuring 
dissipative behavior.

\end{document}